\documentclass[aps,pra,twocolumn,floatfix,groupedaddress,10pt]{revtex4-2}

\usepackage{amsmath,amsthm,amssymb,amscd}
\usepackage{bm}
\usepackage{here}
\usepackage{color}
\usepackage[dvips]{graphicx}
\newtheorem*{theorem}{Theorem}

\begin{document}


\title{Linear simultaneous measurements of position and momentum with minimum error-trade-off
in each minimum uncertainty state}


\author{Kazuya Okamura}
\email[]{k.okamura.renormalizable@gmail.com}
\affiliation{Research Origin for Dressed Photon, 3-13-19 Moriya-cho, Kanagawa-ku, Yokohama, Kanagawa 221-0022, Japan}
\affiliation{Graduate School of Informatics, Nagoya University, Chikusa-ku, Nagoya 464-8601, Japan}


\date{\today}

\begin{abstract}
So-called quantum limits and their achievement are important themes in physics.
Heisenberg's uncertainty relations are the most famous of them
but are not universally valid and violated in general.
In recent years, the reformulation of uncertainty relations is actively studied,
and several universally valid uncertainty relations are derived.
On the other hand, several measuring models, in particular, spin-1/2 measurements, are constructed and
quantitatively examined.
However, there are not so many studies on simultaneous measurements of position and momentum despite their importance.
Here we show that an error-trade-off relation (ETR), called the Branciard-Ozawa ETR,
for simultaneous measurements of position and momentum
gives the achievable bound in minimum uncertainty states.
We construct linear simultaneous measurements of position and momentum that
achieve the bound of the Branciard-Ozawa ETR in each minimum uncertainty state.
To check their performance, we then calculate probability distributions
and families of posterior states, sets of states after the measurements, when using them.
The results of the paper show the possibility of developing
the theory of simultaneous measurements of incompatible observables.
In the future, it will be widely applied to quantum information processing.
\end{abstract}


\maketitle
\section{Introduction \label{1}}
In quantum physics, uncertainty relations and construction of measurement models are
important themes since Heisenberg \cite{Heisenberg1927} and von Neumann \cite{von2018mathematical}.
In the last forty years, quantum measurement theory has developed.
There has been a great deal of study of quantum measurement focused on applications
to quantum information technology nowadays.
Above all, the theory of uncertainty relations
\cite{ozawa2003universally,ozawa2003physical,ozawa2004uncertainty2,ozawa2004uncertainty,
hall2004prior,busch2007heisenberg,lund2010measuring,
branciard2013error,branciard2014deriving,ozawa2014errordisturbance2,
busch2013proof,busch2014colloquium,busch2014measurement,ozawa2013disproving,buscemi2014noise,
dressel2014certainty,korzekwa2014operational}, the central topic of the paper,
has advanced dramatically in the last two decades. Experimental tests of uncertainty relations 
\cite{erhart2012experimental,sulyok2013violation,baek2013experimental,
kaneda2014experimental,ringbauer2014experimental,sulyok2015experimental,demirel2016experimental,
demirel2019experimental,liu2019experimental,liu2019experimental2}
also have been performed due to the rapid improvement of experimental techniques in recent years.
In the paper, we present linear simultaneous measurements
of position and momentum with minimum error-trade-off in each minimum uncertainty state.
The construction of measurements of observables with minimum uncertainty in some class of states
is significant but there are few examples.
In fact, such measurements are given for spin \cite{baek2013experimental,ozawa2014errordisturbance2}
and position \cite{okamura2020linear}.
Therefore, we believe that the results of the paper are an important contribution.

Here we consider a one-dimensional nonrelativistic single-particle system $\mathbf{S}$
 whose position $Q_1$ and momentum $P_1$
are defined as self-adjoint operators on $\mathcal{H}_\mathbf{S}=L^2(\mathbb{R})$ and satisfy the canonical commutation relation
$[Q_1,P_1]=i\hbar 1$. A unit vector $\psi$ in $\mathcal{H}_\mathbf{S}$ is called a minimum uncertainty state
if it satisfies $\sigma(Q_1\Vert\psi)\sigma(P_1\Vert\psi)=\hbar/2$.
\textit{Throughout the paper, we suppose that the state $\psi$ of $\mathbf{S}$
is a minimum uncertainty state with $\langle Q_1 \rangle_\psi=q_1$, $\langle P_1 \rangle_\psi=p_1$
and $\sigma(Q_1\Vert\psi)=\sigma_1$, i.e.,}
\begin{equation}
\psi(x)=\sqrt[4]{\dfrac{1}{(2\pi)\sigma_1^2}}e^{-\frac{(x-q_1)^2}{4\sigma_1^2}+i\frac{p_1}{\hbar}x}
\end{equation}
\textit{in the coordinate representation.}
Minimum uncertainty states appear in Heisenberg's original paper \cite{Heisenberg1927}
and are also called Gaussian wave packets.

In order to define linear simultaneous measurements of $Q_1$ and $P_1$,
we prepare a probe system $\mathbf{P}$ whose positions $Q_2,Q_3$ and momenta $P_2,P_3$
 are described by self-adjoint operators on $\mathcal{H}_\mathbf{P}=L^2(\mathbb{R}^2)$
 and satisfy $[Q_2,Q_3]=[P_2,P_3]=0$ and $[Q_j,P_k]=i\hbar \delta_{jk} 1$ for $j,k=2,3$,
and whose states are described by density operators on $\mathcal{H}_\mathbf{P}$.
$\mathbf{P}$ is supposed to be a one-dimensional nonrelativistic two-particle system or
a two-dimensional nonrelativistic single-particle system.
$Q_2$ and $P_3$ are used as the meters to measure $Q_1$ and $P_1$, respectively. 
\textit{In considering linear simultaneous measurements of position and momentum from now on,
we ignore the intrinsic dynamics of $\mathbf{S}$ and $\mathbf{P}$.}
Here we adopt the following interaction Hamiltonian,
the measurement interaction between $\mathbf{S}$ and $\mathbf{P}$:
\begin{eqnarray}
H_{int} =K[\alpha_1 Q_1P_2+\beta_1 P_1Q_2+\gamma_1(Q_1P_1-Q_2P_2)  \nonumber\\
 +\alpha_2 Q_2P_3+\beta_2 P_2Q_3+\gamma_2(Q_2P_2-Q_3P_3) \nonumber \\
 +\alpha_3 Q_3P_1+\beta_3 P_3Q_1+\gamma_3(Q_3P_3-Q_1P_1)],
\label{eq:one}
\end{eqnarray}
where $K$ is a positive real number, the coupling constant,
and $\alpha_1$, $\alpha_2$, $\alpha_3$, $\beta_1$, $\beta_2$, $\beta_3$, $\gamma_1$, $\gamma_2$
and $\gamma_3$ are real numbers.
This interaction is a natural extension of linear measurements given by Ozawa \cite{ozawa1990quantum}
to simultaneous measurements.
His model is exactly solvable and
contains both the error-free linear position measurement \cite{ozawa1988measurement}
and von Neumann's model \cite{von2018mathematical}.
In particular, the former contributed to the resolution of the dispute on
the sensitivity limit to the gravitational wave detector
(see also \cite{caves1980measurement,yuen1983contractive,caves1985defense,ozawa1989realization,maddox1988beating}).

We treat an error-trade-off relation (ETR) based on the noise-operator based q-rms error $\varepsilon(A)$
for each observable $A$. This error is considered standard and is defined later.
For every simultaneous measurement of $Q_1$ and $P_1$,
the errors $\varepsilon(Q_1)$ of $Q_1$ and $\varepsilon(P_1)$ of $P_1$ in $\psi$ then satisfy
\begin{equation}
\varepsilon(Q_1)^2\sigma(P_1)^2+\sigma(Q_1)^2\varepsilon(P_1)^2\geq \hbar^2/4, \label{BOQP}
\end{equation}
which is a special case of the Branciard-Ozawa ETR.
We say that a simultaneous measurement of $Q_1$ and $P_1$ has the minimum error-trade-off in $\psi$
if it achieves the lower bound of Eq.(\ref{BOQP}) in $\psi$, that is to say,
it satisfies
\begin{equation}
\varepsilon(Q_1)^2\sigma(P_1)^2+\sigma(Q_1)^2\varepsilon(P_1)^2= \hbar^2/4 \label{BOQP2}
\end{equation}
in $\psi$.
As suggested by the existence of the error-free linear position measurements,
Heisenberg's ETR, one of his uncertainty relations,
\begin{equation}
\varepsilon(Q_1)\varepsilon(P_1)\geq \hbar/2 \label{HETR}
\end{equation}
is violated in general. Its violation always occurs when we use
linear simultaneous measurements of $Q_1$ and $P_1$ with the minimum error-trade-off in each minimum uncertainty state.
A famous example of simultaneous measurement of position and momentum
is the Arthurs-Kelly model (see \cite{arthurs1965bstj} and Methods).
Since their model is motivated by von Neumann's model
and satisfies Heisenberg's ETR, it has been considered plausible.
On the other hand, our discussion is based on the general description of measuring processes
in modern quantum measurement theory.
The general theory of quantum measurement tells us that a broader class of simultaneous measurement models 
besides the Arthurs-Kelly model is physically valid.
We expect that our models introduced in the paper become the new, good example.

In Sec.~\ref{2}, measuring process and the noise-operator based q-rms error are defined.
Linear simultaneous measurement of position and momentum is then defined.
In Sec.~\ref{3}, we first present a theorem that gives a necessary and sufficient condition 
for a linear simultaneous measurement of position and momentum to satisfy Eq.~(\ref{BOQP2}) in $\psi$.
Next, we give four families of linear simultaneous measurements of position and momentum 
which satisfy Eq.~(\ref{BOQP2}) in $\psi$.
We then investigate probability distributions and states after the measurement when using such families
of linear simultaneous measurements of position and momentum.
In Sec.~\ref{4}, the results of the paper are examined.
In Sec.~\ref{5}, we prove the theorem and show a systematic construction of 
linear simultaneous measurements of position and momentum which satisfy Eq.~(\ref{BOQP2}) in $\psi$.

\textbf{Conventions.}
Let $\mathcal{H}$ be a Hilbert space.
For every self-adjoint operator $X$ on $\mathcal{H}$, $E^X$ denotes its spectral measure.
Let $n$ be a natural number, and $X_1,\cdots,X_n$ mutually commuting self-adjoint operators on $\mathcal{H}$,
and $\phi$ a unit vector in $\mathcal{H}$.
The expectation value and standard deviation of an observable $X$ in a vector state $\phi$ are denoted by
\begin{eqnarray}
\langle X \rangle=\langle X \rangle_\phi=\langle\phi| X \phi\rangle=\langle\phi| X |\phi\rangle,\\
\sigma(X)=\sigma(X\Vert\phi)=\langle \phi| (X- \langle X \rangle_\phi)^2 \phi\rangle^{\frac{1}{2}},
\label{eq}
\end{eqnarray}
respectively.
Then the (joint) probability measure $\mu_\phi^{X_1,\cdots,X_n}$ of $X_1,\cdots,X_n$ in $\phi$ is defined by
\begin{equation}
\mu_\phi^{X_1,\cdots,X_n}(I_1\times\cdots\times I_n)
=\langle \phi|E^{X_1}(I_1)\cdots E^{X_n}(I_n)\phi \rangle
\end{equation}
for all intervals(, more generally, all Borel sets) $I_1,\cdots, I_n$ of $\mathbb{R}$.
$p_\phi^{X_1,\cdots,X_n}(x_1,\cdots,x_n)$ denotes the probability density function of
$\mu_\phi^{X_1,\cdots,X_n}$ with respect to the Lebesgue measure on $\mathbb{R}^n$ if it exists.
For every linear operator $X$ and $Y$ on $\mathcal{H}$ and $\mathcal{K}$,
linear operators $X\otimes Y$, $X\otimes 1$ and $1\otimes Y$ on $\mathcal{H}\otimes\mathcal{K}$ are
abbreviated as $XY$, $X$ and $Y$, respectively.

\section{Preliminaries \label{2}}

\subsection{Measuring process\label{2.1}}

First, we shall define a measuring process for $\mathbf{S}$,
which is a quantum mechanical modeling of the probe part $\mathbf{P}_0$ of a measuring apparatus $\mathbf{A}_0$.
Let $n$ be a natural number.
Here a $(n+3)$-tuple $\mathbb{M}_0=(\mathcal{K},\zeta,M_1,\cdots,M_n,U)$ is called
a $n$-meter measuring process for $\mathbf{S}$ (or for $\mathcal{H}_\mathbf{S}$)
if it satisfies the following conditions:
$(1)$ $\mathcal{K}$ is a Hilbert space.
$(2)$ $\zeta$ is a unit vector of $\mathcal{K}$, the vector state of $\mathbf{P}_0$,
$(3)$ $M_1,\cdots,M_n$ are mutually commuting self-adjoint operators on $\mathcal{K}$ as meters,
mutually compatible observables of $\mathbf{P}_0$,
$(4)$ $U$ is a unitary operator on $\mathcal{H}_\mathbf{S}\otimes\mathcal{K}$,
the measuring interaction which turns on at time $0$
and turns off at time $\tau$ between $\mathbf{S}$ and $\mathbf{P}_0$.
We then adopt the following notation for every linear operator $Z$
on $\mathcal{H}_\mathbf{S}\otimes\mathcal{K}$:
\begin{equation}
Z(0)=Z,\hspace{5mm}Z(\tau)= U^\dag ZU. \label{Eq1}
\end{equation}
A $2$-meter measuring process $(\mathcal{K},\zeta,M_1,M_2,U)$ for $\mathbf{S}$
is called a simultaneous measurement of position $Q_1$ and momentum $P_1$ or
a simultaneous $(Q_1,P_1)$-measurement if $M_1$ and $M_2$ are used to measure $Q_1$ and $P_1$, respectively.

Let $n$ be a natural number.
Let $X_1,\cdots,X_n$ be observables of $\mathbf{S}$, $\phi$ a vector state of $\mathbf{S}$,
and $\mathbb{M}_0=(\mathcal{K},\zeta,M_1,\cdots,M_n,U)$
a $n$-meter measuring process for $\mathbf{S}$.
We consider that $X_1,\cdots,X_n$ are measured in terms of $\mathbb{M}_0=(\mathcal{K},\zeta,M_1,\cdots,M_n,U)$,
and that $X_1,\cdots,X_n$ are compared with $M_1,\cdots,M_n$, respectively.
The noise-operator based q-rms error $\varepsilon(X_j)=\varepsilon(X_j,\mathbb{M}_0,\phi)$
of $X_j$ is then defined by
\begin{equation}
\varepsilon(X_j)=
\varepsilon(X_j,\mathbb{M}_0,\phi)=\langle N_j^2 \rangle_{\phi\otimes \zeta}^{\frac{1}{2}}
\end{equation}
for all $j=1,\cdots,n$, where $N_j$ is the noise operator defined by
\begin{equation}
N_j=N(X_j,\mathbb{M}_0)=M_j(\tau)-X_j(0)
\end{equation}
for all $j=1,\cdots,n$.
The error defined here is applicable to the case where $X_j(0)$ and $M_j(\tau)$ does not commute,
and is considered standard.

For every simultaneous $(Q_1,P_1)$-measurement $\mathbb{M}_0=(\mathcal{K},\zeta,M_1,M_2,U)$,
Eq.~(\ref{BOQP}) holds in $\psi$ for 
\begin{align}
\varepsilon(Q_1) &= \varepsilon(Q_1,\mathbb{M}_0,\psi)
=\langle (M_1(\tau)-Q_1(0))^2 \rangle_{\psi\otimes \zeta}^{\frac{1}{2}},  \\
\varepsilon(P_1) &= \varepsilon(P_1,\mathbb{M}_0,\psi)=
\langle (M_2(\tau)-P_1(0))^2 \rangle_{\psi\otimes \zeta}^{\frac{1}{2}}.
\end{align}

\subsection{Linear simultaneous measurement of position and momentum\label{2.2}}

A $2$-meter measuring process $\mathbb{M}=(\mathcal{H}_\mathbf{P},\xi,Q_2,P_3,U(\tau))$ for $\mathbf{S}$ is called
a linear simultaneous measurement of position $Q_1$ and momentum $P_1$
or a linear simultaneous $(Q_1,P_1)$-measurement
if $Q_2$ and $P_3$ are used to measure $Q_1$ and $P_1$, respectively, where 
$\xi$ is a unit vector of $\mathcal{H}_\mathbf{P}=L^2(\mathbb{R}^2)$ satisfying
$\Vert Q_2^{m_2}Q_3^{m_2}P_2^{n_2}P_3^{n_3}\xi\Vert<+\infty$ for all non-negative integers $m_2,m_3,n_2,n_3$,
$\tau(>0)$ is the time the measurement finishes and $U(t)$ is defined by $U(t)=e^{-itH_{int}/\hbar}$
for all $t\in\mathbb{R}$.
\textit{Since we ignore the intrinsic dynamics of $\mathbf{S}$ and $\mathbf{P}$,
$K$ contributes only to the time scale of the measurement time. For simplicity, we assume $K=1$ in the paper.}
For every observable $Z$ of $\mathbf{S}+\mathbf{P}$ at time $0$ and $t\in\mathbb{R}$,
the same observable $Z(t)$ at time $t$ is given by
\begin{equation}
Z(t)=U(t)^\dag Z U(t)
\end{equation}
for all $t\in\mathbb{R}$.
This is consistent with the notation before, Eq.(\ref{Eq1}).
By solving Heisenberg's equations of motion, we have
\begin{eqnarray}
\left(
\begin{array}{c}
Q_1(t)  \\
Q_2(t)  \\
Q_3(t) 
\end{array}
\right)=
e^{tR}\left(
\begin{array}{c}
Q_1(0)  \\
Q_2(0)  \\
Q_3(0) 
\end{array}
\right), \label{HEM1} \\
\left(
\begin{array}{c}
P_1(t)  \\
P_2(t)  \\
P_3(t) 
\end{array}
\right)=
e^{-tR^T}\left(
\begin{array}{c}
P_1(0)  \\
P_2(0)  \\
P_3(0) 
\end{array}
\right) \label{HEM2}
\end{eqnarray}
for all $t\in\mathbb{R}$, where
\begin{equation}
R=\left(
\begin{array}{ccc}
\gamma_1-\gamma_3 & \beta_1& \alpha_3 \\
\alpha_1 &\gamma_2-\gamma_1 & \beta_2 \\
\beta_3 & \alpha_2 & \gamma_3-\gamma_2
\end{array}
\right)
\end{equation}
and $R^T$ denotes the transpose of $R$.
We see that $e^{tR},e^{-tR^T}\in SL(3,\mathbb{R})$ for all $t\in\mathbb{R}$.
$e^{\tau R}$ and $e^{-\tau R^T}$ are denoted by $A=(a_{ij})$ and $B=(b_{ij})$, respectively.
When we use a linear simultaneous $(Q_1,P_1)$-measurement,
the noise-operator based q-rms errors $\varepsilon(Q_1)$ and $\varepsilon(P_1)$ have
the following representations:
\begin{align}
\varepsilon(Q_1)^2 &=\varepsilon(Q_1,\mathbb{M},\psi)^2 \nonumber\\
&=(a_{21}-1)^2\sigma(Q_1\Vert\psi)^2+\sigma(a_{22}Q_2+a_{23}Q_3\Vert\xi)^2 \nonumber\\
 &\hspace{2mm}+((a_{21}-1)\langle Q_1 \rangle_\psi+a_{22}\langle Q_2\rangle_\xi+a_{23}\langle Q_3\rangle_\xi)^2,  \\
\varepsilon(P_1)^2 &=\varepsilon(P_1,\mathbb{M},\psi)^2 \nonumber\\
&=(b_{31}-1)^2\sigma(P_1\Vert\psi)^2+\sigma(b_{32}P_2+b_{33}P_3\Vert\xi)^2 \nonumber \\
 &\hspace{5mm}+((b_{31}-1)\langle Q_1 \rangle_\psi+b_{32}\langle P_2\rangle_\xi+b_{33}\langle P_3\rangle_\xi)^2. 
\end{align}

\section{Results \label{3}}
\subsection{Characterization theorem\label{3.1}}

\begin{figure}[H]
\begin{center}
\includegraphics[clip]{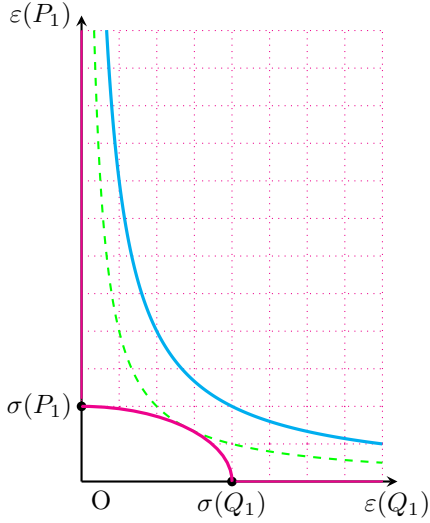}
\caption{When the state of $\mathbf{S}$ is $\psi$,
possible values of the pair $(\varepsilon(Q_1),\varepsilon(P_1))$ of the errors
are indicated by the area with a grid of dotted magenta lines
and with magenta boundary
except for two points $(\sigma(Q_1),0)$ and $(0,\sigma(P_1))$. By the theorem, 
$\varepsilon(Q_1)^2\sigma(P_1)^2+\sigma(Q_1)^2\varepsilon(P_1)^2=\hbar^2/4$
($\varepsilon(Q_1),\varepsilon(P_1)>0$), a part of its boundary,
is achieved by linear simultaneous $(Q_1,P_1)$-measurements,
and gives the unbreakable limitation for the pair $(\varepsilon(Q_1),\varepsilon(P_1))$.
The cyan line is Heisenberg's bound, $\varepsilon(Q_1)\varepsilon(P_1)=\hbar/2$.
On the other hand, the dashed green line indicates $\varepsilon(Q_1)\varepsilon(P_1)=\hbar/4$.}\label{fig1}
\end{center}
\end{figure}

The following theorem is the first result of the paper:
\begin{theorem}{}\label{Main2}
A linear simultaneous $(Q_1,P_1)$-measurement
$\mathbb{M}=(\mathcal{H}_\mathbf{P},\xi,Q_2,P_3,U(\tau))$ satisfies Eq.~(\ref{BOQP2})
in $\psi$ if and only if it satisfies the following three conditions:\\
$(i)$ $\displaystyle{(a_{21}-1)\langle Q_1 \rangle_\psi
+a_{22}\langle Q_2 \rangle_\xi+a_{23}\langle Q_3 \rangle_\xi=0}$
 and $(b_{31}-1)\langle P_1 \rangle_\psi+b_{32}\langle P_2 \rangle_\xi+b_{33}\langle P_3 \rangle_\xi=0$.\\
$(ii)$ $\sigma(a_{22}Q_2+a_{23}Q_3\Vert\xi) = |a_{21}b_{31}|^{\frac{1}{2}}\sigma(Q_1\Vert\psi)$ and
$\sigma(b_{32}P_2+b_{33}P_3\Vert\xi) = |a_{21}b_{31}|^{\frac{1}{2}}\sigma(P_1\Vert\psi)$.\\
$(iii)$ $a_{21}>0$, $b_{31}>0$ and $a_{21}+b_{31}=1$.

Furthermore, for every $\nu\in(0,1)$, there exists a linear simultaneous $(Q_1,P_1)$-measurement
such that
\begin{equation}\label{MUET}
\varepsilon(Q_1)^2=(1-\nu)\sigma(Q_1)^2\hspace{3mm}\text{and}\hspace{3mm}\varepsilon(P_1)^2=\nu\sigma(P_1)^2
\end{equation}
in $\psi$.
\end{theorem}
By the above theorem, any linear simultaneous $(Q_1,P_1)$-measurement with the minimum error-trade-off in $\psi$ satisfies
\begin{equation}
\varepsilon(Q_1)<\sigma(Q_1),\hspace{5mm}\varepsilon(P_1)<\sigma(P_1),
\end{equation}
and
\begin{equation}
\varepsilon(Q_1)\varepsilon(P_1)=\dfrac{\hbar}{2}\sqrt{\dfrac{1}{4}-\left(\nu-\dfrac{1}{2} \right)^2}
\leq \dfrac{\hbar}{4} <\dfrac{\hbar}{2}.
\end{equation}
Thus, the range of possible values of the error pairs
$(\varepsilon(Q_1),\varepsilon(P_1))$ in the state $\psi$ is as shown in FIG. \ref{fig1}.

\subsection{Concrete models\label{3.2}}
The above theorem does not directly tell us how to construct
simultaneous $(Q_1,P_1)$-measurements with the minimum error-trade-off in $\psi$.
Notably, in contrast to exactly solvable linear measurements \cite{ozawa1990quantum},
$e^{tR}$ has no more explicit formula. Therefore, we adandon analyzing $e^{tR}$ as it is.
\textit{We remind the reader that $K=1$ is assumed.}
We shall give a novel, exactly solvable subclass of linear simultaneous $(Q_1,P_1)$-measurements.
The following two constraints for $R$ are imposed:\\
$(\mathrm{C1})$ $\alpha_2=\beta_2=\gamma_1=\gamma_3=0$.\hspace{3mm}
 $(\mathrm{C2})$ $\alpha_1\beta_1=\alpha_3\beta_3$.\\
Under these constraints, $R$ is denoted by $S$, that is,
\begin{equation}
S=\left(
\begin{array}{ccc}
0 & \beta_1 & \alpha_3 \\
\alpha_1 & \gamma_2 & 0 \\
\beta_3 & 0 & -\gamma_2
\end{array}
\right).
\end{equation}

Let $\nu\in(0,1)$ and $\kappa\in\mathbb{R}\backslash \{0\}$. We define
a state $\xi_{\nu,\kappa}$ of $\mathbf{P}$, which satisfies the following conditions:
$(1)$ $\sigma(Q_2)\sigma(P_2)=\sigma(Q_3)\sigma(P_3)=\hbar/2$ and
$\langle Q_2Q_3 \rangle=\langle Q_2 \rangle\langle Q_3 \rangle$,
$(2)$ $\sigma(Q_2)=\sqrt{\frac{\nu(1-\nu)}{2\kappa^2}}\sigma_1$ and
$\sigma(Q_3)=\sqrt{\frac{2\kappa^2}{\nu(1-\nu)}}\sigma_1$,
$(3)$ $\langle Q_2 \rangle=\frac{1-\nu}{\kappa}q_1$, 
$\langle Q_3 \rangle=0$, 
$\langle P_2 \rangle=0$ and
$\langle P_3 \rangle=\frac{\nu}{\kappa}p_1$,
i.e.,
\begin{equation}
\xi_{\nu,\kappa}(x_{2,3})=\dfrac{1}{\sqrt{(2\pi)\sigma_1^2}}
e^{-\frac{1}{4}\Vert G^{-\frac{1}{2}} (x_{2,3}-u)\Vert^2+\frac{i}{\hbar}\langle v,x_{2,3} \rangle}
\end{equation}
for all $x_{2,3}=\left(
\begin{array}{c}
x_2  \\
x_3
\end{array}
\right)\in\mathbb{R}^2$ in the coordinate representation, where
$u=\left(
\begin{array}{c}
\frac{1-\nu}{\kappa}q_1 \\
0 
\end{array}
\right)$, $v=\left(
\begin{array}{c}
0  \\
\frac{\nu}{\kappa}p_1
\end{array}
\right)$ and
$G= \sigma_1^2\left(
\begin{array}{cc}
\frac{\nu(1-\nu)}{2\kappa^2} & 0 \\
0 & \frac{2\kappa^2}{\nu(1-\nu)}
\end{array}
\right)$.

For every $\nu\in(0,1)$, we present four linear simultaneous $(Q_1,P_1)$-measurements
$(\mathcal{H}_\mathbf{P},\xi_{\nu,\kappa},Q_2,P_3,U(\tau))$
satisfying Eq.~(\ref{MUET}) herein,
denoted by $\mathbb{X}_\nu$, $\mathbb{Y}_\nu^2$, $\mathbb{Y}_\nu^0$ and $\mathbb{Z}_\nu$, respectively.
Each model is specified by the triplet of $\tau$, $S$ and $\kappa$ in the following table, Table \ref{t1}.
\begin{table}[H] 
\begin{center}
\caption{}
\begin{ruledtabular}
\begin{tabular}{ccccc} 
 & $\tau$ & $S$ & $\kappa$ & $E$ \\ \hline 
$\mathbb{X}_\nu$ & $\dfrac{\pi}{2}$ & $\left(
\begin{array}{ccc}
0 &-\frac{2}{\nu} &-\frac{1-\nu}{2}  \\
\frac{\nu}{2} &1 &0  \\
\frac{2}{1-\nu} & 0& -1
\end{array}
\right)$ & $2$ & $1$ \\ \hline
$\mathbb{Y}_\nu^2$ & 1 & $\left(
\begin{array}{ccc}
0 & -\frac{4}{\nu}& \frac{\nu-1}{2} \\
\frac{\nu}{2} &2 & 0 \\
\frac{4}{1-\nu} & 0& -2
\end{array}
\right)$ & $4$ & $0$ \\ \hline
$\mathbb{Y}_\nu^0$ & 1 & $\left(
\begin{array}{ccc}
0 & 0 & -(1-\nu) \\
\nu &0 & 0 \\
0 & 0& 0
\end{array}
\right)$ & $1$ & $0$ \\ \hline
$\mathbb{Z}_\nu$ & $\log 2$  & $\left(
\begin{array}{ccc}
0 &0 & \nu-1  \\
\nu &1 & 0 \\
0 & 0& -1
\end{array}
\right)$ & $2$ & $-1$ \\ 
\end{tabular} \label{t1}
\end{ruledtabular}
\end{center}
\end{table}
Here $E$ is a real number defined by
\begin{equation}
\alpha_1\beta_1=\alpha_3\beta_3=-\dfrac{\gamma_2^2+E}{2},
\end{equation}
and is used to explicitly solve $e^{tS}$ for all $t\in\mathbb{R}$ (see Sec.~\ref{5.1}).

\subsection{Probability distributions and families of posterior states\label{3.3}}

Our next interest is to give probability distributions and families of posterior states when using
concrete models $\mathbb{X}_\nu$, $\mathbb{Y}_\nu^2$, $\mathbb{Y}_\nu^0$ and $\mathbb{Z}_\nu$.
First, we show probability distributions related to $\varepsilon(Q_1)$ and $\eta(P_1)$,
and check the validity of $\mathbb{X}_\nu$, $\mathbb{Y}_\nu^2$, $\mathbb{Y}_\nu^0$ and $\mathbb{Z}_\nu$.
For every $\nu\in(0,1)$, whether we use $\mathbb{X}_\nu$, $\mathbb{Y}_\nu^2$, $\mathbb{Y}_\nu^0$ or $\mathbb{Z}_\nu$,
we get the following probability density functions:
\begin{align}
p^{Q_2(\tau),P_3(\tau)}_{\psi\otimes\xi_{\nu,\kappa}}(z,w)
&= p_{\nu\sigma_1^2}(z-q_1)p_{(1-\nu)\hat{\sigma}_1^2}(w-p_1), \label{JDQ1}\\
p^{Q_1(0),Q_2(\tau)}_{\psi\otimes\xi_{\nu,\kappa}}(x,z)
&=p_{(1-\nu)\sigma_1^2}(x-z)p_{\nu\sigma_1^2}(z-q_1), \label{JDQ2}\\
p^{P_1(0),P_3(\tau)}_{\psi\otimes\xi_{\nu,\kappa}}(y,w) &= 
p_{\nu\hat{\sigma}_1^2}(y-w)p_{(1-\nu)\hat{\sigma}_1^2}(w-p_1), \label{JDQ3}
\end{align}
where $\hat{\sigma}_1=\hbar/(2\sigma_1)$ and
$p_{\sigma^2}(x)$ denotes the probability density function of the Gaussian probability measure
with mean $0$ and variance $\sigma^2$ (equivalently, standard deviation $\sigma$), i.e.,
\begin{equation}
p_{\sigma^2}(x)=\dfrac{1}{\sqrt{(2\pi)\sigma^2}}e^{-\frac{1}{2\sigma^2}x^2}.
\end{equation}
We see that all of Eqs.~(\ref{JDQ1}), (\ref{JDQ2}) and (\ref{JDQ3}) depend on $\psi$ and $0<\nu<1$.
Of the three equations, only Eq.~(\ref{JDQ1}) can be directly confirmed
by any of $\mathbb{X}_\nu$, $\mathbb{Y}_\nu^2$, $\mathbb{Y}_\nu^0$ or $\mathbb{Z}_\nu$.
The rest two equations, Eqs.~(\ref{JDQ2}) and (\ref{JDQ3}), are essential
for understanding the performance of $\mathbb{X}_\nu$, $\mathbb{Y}_\nu^2$, $\mathbb{Y}_\nu^0$ and $\mathbb{Z}_\nu$.
From Eq.~(\ref{JDQ2}), the probability density function of the conditional probability measure of $Q_1(0)$
in $\psi\otimes\xi_{\nu,\kappa}$  under the condition that the value $z$ of $Q_2(\tau)$ is given is determined as
\begin{equation}
p^{Q_1(0)}_{Q_2(\tau)=z,\psi\otimes\xi_{\nu,\kappa}}(x)=p_{(1-\nu)\sigma_1^2}(x-z). \label{CPM100}
\end{equation}
Since $\varepsilon(Q_1)^2=(1-\nu)\sigma_1^2$, Eq.~(\ref{CPM100}) means that,
when the value $z$ of $Q_2(\tau)$ is output, $Q_1(0)$ obeys
the Gaussian probability measure with mean $z$ and standard deviation $\varepsilon(Q_1)$.
The same argument can be made for Eq.~(\ref{JDQ3}) and $\varepsilon(P_1)^2=\nu\hat{\sigma}_1^2$.
The noise-operator based q-rms errors $\varepsilon(Q_1)$ and $\varepsilon(P_1)$ are then equal
to Gauss' errors $\varepsilon_G(\mu^{Q_1(0),Q_2(\tau)}_{\psi\otimes\xi})$ and 
$\varepsilon_G(\mu^{P_1(0),P_3(\tau)}_{\psi\otimes\xi})$, respectively, i.e.,
\begin{equation}
\varepsilon(Q_1)=\varepsilon_G(\mu^{Q_1(0),Q_2(\tau)}_{\psi\otimes\xi}),\hspace{3mm}
\varepsilon(P_1)=\varepsilon_G(\mu^{P_1(0),P_3(\tau)}_{\psi\otimes\xi}). \label{COR}
\end{equation}
Here Gauss' error $\varepsilon_G(\mu)$ for a probability distribution $\mu$ on $\mathbb{R}^2$ is defined by
\begin{equation}
\varepsilon_G(\mu)=\left(\int_{\mathbb{R}^2} (x-y)^2\;d\mu(x,y)\right)^{\frac{1}{2}}. \label{GE1}
\end{equation}
Following Laplace's pioneering work, Gauss \cite{Gauss1995theory} defined his error in 1821.
His error is now redefined as above and widely used in the setting of measure-theoretical probability theory.

Next, we consider a family of posterior states, which is the set of the states after the measurement
for each output value of the meter (see \cite{ozawa1985,okamura2016measurement} for the general theory).
It is difficult to find families of posterior states for general
linear simultaneous $(Q_1,P_1)$-measurements with the minimum error-trade-off in $\psi$.
Here we shall give them for $\{\mathbb{Y}^0_\nu\}_{\nu\in(0,1)}$
and $\{\mathbb{Z}_\nu\}_{\nu\in(0,1)}$.
For every $\nu\in(0,1)$, the family $\{\psi_y\}_{y\in\mathbb{R}^2}$
of posterior states for $(\mathbb{Y}^0_\nu,\psi)$ is the set of
the minimum uncertainty state $\psi_y$ with $\langle Q_1 \rangle_{\psi_y}=\frac{y_1-(1-\nu)q_1}{\nu}$,
$\langle P_1 \rangle_{\psi_y}=\frac{y_2-\nu p_1}{1-\nu}$
and $\sigma(Q_1\Vert\psi_y)=\sqrt{\frac{1-\nu}{\nu}}\sigma_1$ for all $y=\left(
\begin{array}{c}
y_1  \\
 y_2
\end{array}
\right)\in\mathbb{R}^2$, i.e.,
\begin{equation}
\psi_y(x)=\frac{e^{-\frac{\nu}{4(1-\nu)\sigma_1^2}\left(x-\frac{y_1-(1-\nu)q_1}{\nu} \right)^2
+i\frac{y_2-\nu p_1}{(1-\nu)\hbar}x}}{\sqrt[4]{\frac{2\pi(1-\nu)\sigma_1^2}{\nu}}} \label{FPSY}
\end{equation}
for all $y=\left(
\begin{array}{c}
y_1  \\
 y_2
\end{array}
\right)\in\mathbb{R}^2$ in the coordinate represenation. 
For every $\nu\in(0,1)$, the family $\{\psi_y\}_{y\in\mathbb{R}^2}$
of posterior states for $(\mathbb{Z}_\nu,\psi)$ is the same as that for $(\mathbb{Y}^0_\nu,\psi)$.

\section{Discussion\label{4}}

\subsection{The Arthurs-Kelly model\label{4.1}}

Here we shall mention the differences between this paper
and the paper \cite{arthurs1965bstj} of Arthurs and Kelly, an important previous study,
on the treatment of simultaneous measurements of position and momentum.
They use the
$2$-meter measuring process $\mathbb{M}_{\mathrm{AK}}=(\mathcal{H}_\mathbf{P},\xi,Q_2,Q_3,U_{\mathrm{AK}}(K^{-1}))$
for $\mathbf{S}$, where $U_{\mathrm{AK}}(t)=e^{-itH_{\mathrm{AK}}/\hbar}$ is
a one-parameter group on
$\mathcal{H}_\mathbf{S}\otimes\mathcal{H}_\mathbf{P}=L^2(\mathbb{R})\otimes L^2(\mathbb{R}^2)$
with $H_{\mathrm{AK}}=K(Q_1P_2+P_1P_3)$,
and use $Q_2$ and $Q_3$ to measure $Q_1$ and $P_1$, respectively.
Their interaction Hamiltonian is obtained from that of the linear simultaneous $(Q_1,P_1)$-measurement
with $\alpha_1=-\alpha_3=1$ and $\alpha_2=\beta_1=\beta_2=\beta_3=\gamma_1=\gamma_2=\gamma_3=0$
by replacing $Q_3$ and $P_3$ by $-P_3$ and $Q_3$, respectively. Then we have
\begin{eqnarray}
U_{\mathrm{AK}}(K^{-1})^\dag Q_2 U_{\mathrm{AK}}(K^{-1})=Q_1+Q_2+\frac{1}{2}P_3,\\
U_{\mathrm{AK}}(K^{-1})^\dag Q_3 U_{\mathrm{AK}}(K^{-1})=P_1-\frac{1}{2}P_2+Q_3,
\end{eqnarray}
so that the q-rms errors $\varepsilon(Q_1)=\varepsilon(Q_1,\mathbb{M}_{\mathrm{AK}},\psi)$
and $\varepsilon(P_1)=\varepsilon(P_1,\mathbb{M}_{\mathrm{AK}},\psi)$ satisfy Heisenberg's ETR, Eq.~(\ref{HETR}).
This result shows that the Arthurs-Kelly model is not what we desire.

On the other hand, the measuring interaction of Ozawa's exactly solvable linear measurements is given by
\begin{equation}
H_{O} =K[\alpha Q_1P_2+\beta P_1Q_2+\gamma(Q_1P_1-Q_2P_2)],
\end{equation}
where $K$ is a positive real number, the coupling constant,
and $\alpha$, $\beta$ and $\gamma$ are real numbers.
In \cite{ozawa1990quantum}, Ozawa systematically analyzed his exactly solvable measuring models
using this interaction, and calculated the noise-operator baed q-rms error
and the disturbance-operator based q-rms disturbance.
His investigation motivated the author just as von Neumann's work inspired Arthurs and Kelly.

\subsection{The Branciard-Ozawa ETR and the noise-operator based q-rms error\label{4.2}}

The reformulation of uncertainty relations is a currently developing project.
As part of this research project, this study has the significance of
connecting the recent knowledge about uncertainty relations with the construction of measurement models.
After Ozawa's inequality
\begin{equation}
\varepsilon(X)\varepsilon(Y)+\varepsilon(X)\sigma(Y)+\sigma(X)\varepsilon(Y)\geq C_{XY}
\end{equation}
was proved, the study of uncertainty relations became active,
where $C_{XY}=|\mathrm{Tr}(\rho[X,Y])|/2$ and $\rho$ is a density operator on $L^2(\mathbb{R})$
describing the state of $\mathbf{S}$.
Note, however, that the noise-operator based q-rms error $\varepsilon(X)$ and
the standard deviation $\sigma(Y)$ are defined for $\rho$.
The tightest ETR, which is now known, is the Branciard-Ozawa ETR
\begin{align}
 &\hspace{3mm} \varepsilon(X)^2\sigma(Y)^2+\sigma(X)^2\varepsilon(Y)^2\nonumber\\
 &+2\varepsilon(X)\varepsilon(Y)\sqrt{\sigma(X)^2\sigma(Y)^2-D_{XY}^2}\geq D_{XY}^2,
\end{align}
where $D_{XY}=\mathrm{Tr}|\sqrt{\rho}[X,Y]\sqrt{\rho}|/2$ 
satisfies $D_{XY}\geq C_{XY}$ (see \cite{ozawa2014errordisturbance2}). 
This inequality is first proved for pure (vector) states by Branciard \cite{branciard2013error},
and is extended to mixed states by Ozawa \cite{ozawa2014errordisturbance2}.
Eq.~(\ref{BOQP}) is the case where
$X=Q_1$, $Y=P_1$ and the state of $\mathbf{S}$ is $\psi$.

There is a claim that the use of the noise-operator based q-rms error is questionable because
it sometimes vanishes for inaccurate measurements of observables (see \cite{busch2007heisenberg} for example).
In constrast to such a claim, it is shown in \cite{ozawa2019soundness} that the q-rms error satisfies
satisfactory conditions except for the completeness. 
A q-rms error is said to be complete
if it never vanishes for inaccurate measurements of observables in each state \cite{ozawa2019soundness}.
The noise-operator based q-rms error is regarded
as a straightfoward generalization of Gauss' error to quantum measurement.
Instead of sticking to the noise-operator based q-rms error only,
its improved versions that satisfy the completeness are also proposed in \cite{ozawa2019soundness}.
In statistics and information theory, various quantitative measures are defined for different purposes.
In that sense, it is valid that we use the noise-operator based q-rms error as a standard,
and that we use its improved versions as alternatives when its use is problematic.

\section{Methods\label{5}}
As in standard textbooks of quantum mechanics,
$Q_j$ and $P_k$ satisfy
\begin{align}
 (Q_jf)(x_1,x_2,x_3) &= x_jf(x_1,x_2,x_3),\\
(P_kg)(x_1,x_2,x_3) &= \dfrac{\hbar}{i}\dfrac{\partial}{\partial x_k}g(x_1,x_2,x_3), 
\end{align}
respectively, in the coordinate representation for every $j,k=1,2,3$,
and for appropriate functions $f$ and $g$ on $\mathbb{R}^3$.
We do not explicitly use the above representation in the paper.


\subsection{Proof of Theorem and the construction of models\label{5.1}}

To begin with, we shall prove Theorem.
When the state of $\mathbf{S}$ is $\psi$ and
a linear $(Q_1,P_1)$-measurement $\mathbb{M}=(\mathcal{H}_\mathbf{P},\xi,Q_2,P_3,U(\tau))$ is used,
we have the following evaluation:
\begin{align}
 &\hspace{5mm}\varepsilon(Q_1)^2\sigma(P_1)^2+\sigma(Q_1)^2\eta(P_1)^2 \nonumber\\
 &\geq (a_{21}-1)^2\sigma(Q_1)^2\sigma(P_1)^2+\sigma(a_{22}Q_2+a_{23}Q_3)^2\sigma(P_1)^2 \nonumber \\
 &\hspace{5mm}+(b_{31}-1)^2\sigma(Q_1)^2\sigma(P_1)^2+\sigma(Q_1)^2\sigma(b_{32}P_2+b_{33}P_3)^2 \nonumber \\
 &=\dfrac{\hbar^2}{4}\{(a_{21}-1)^2+(b_{31}-1)^2\} \nonumber \\
 &\hspace{5mm} +\sigma(a_{22}Q_2+a_{23}Q_3)^2\sigma(P_1)^2+\sigma(Q_1)^2\sigma(b_{32}P_2+b_{33}P_3)^2\nonumber \\
 &=\dfrac{\hbar^2}{4}\{(a_{21}-1)^2+(b_{31}-1)^2\} \nonumber \\
 &\hspace{5mm}+2\sigma(Q_1)\sigma(P_1)\sigma(a_{22}Q_2+a_{23}Q_3)\sigma(b_{32}P_2+b_{33}P_3) \nonumber \\
 &\hspace{5mm} +(\sigma(a_{22}Q_2+a_{23}Q_3)\sigma(P_1)-\sigma(Q_1)\sigma(b_{32}P_2+b_{33}P_3))^2 \nonumber \\
 &\geq \dfrac{\hbar^2}{4}\{(a_{21}-1)^2+(b_{31}-1)^2\}\nonumber \\
 &\hspace{5mm} +\hbar\sigma(a_{22}Q_2+a_{23}Q_3)\sigma(b_{32}P_2+b_{33}P_3)\nonumber \\
 &\geq \hbar^2 l(a_{21},b_{31}),
\end{align}
where $l(a_{21},b_{31})$ is the function on $\mathbb{R}^2$ defined by
\begin{equation}
l(a_{21},b_{31})=\dfrac{1}{4}\{(a_{21}-1)^2+(b_{31}-1)^2\}+\dfrac{1}{2}|a_{21}b_{31}|,
\end{equation}
and takes the minimal value $1/4$ when $a_{21},b_{31}\geq 0$ and $a_{21}+b_{31}=1$.
By $[Q_2(\tau),P_3(\tau)]=0$, we have $a_{21}b_{31}+a_{22}b_{32}+a_{23}b_{33}=0$.
We see that $a_{22}Q_2+a_{23}Q_3$ and $b_{32}P_2+b_{33}P_3$ satisfy the following commutation relation
\begin{equation}
[a_{22}Q_2+a_{23}Q_3,b_{32}P_2+b_{33}P_3]=i\hbar(-a_{21}b_{31}) 1.
\end{equation}
Therefore, we obtain
\begin{equation}
\sigma(a_{22}Q_2+a_{23}Q_3)\sigma(b_{32}P_2+b_{33}P_3)\geq \dfrac{\hbar}{2}|a_{21}b_{31}|.
\end{equation}

A linear simultaneous $(Q_1,P_1)$-measurement
$\mathbb{M}=(\mathcal{H}_\mathbf{P},\xi,Q_2,P_3,U(\tau))$ satisfies Eq.~(\ref{BOQP2})
in $\psi$ if and only if it satisfies the conditions $(i)$ and\\
$(ii.1)$ $\displaystyle{\sigma(P_1)\sigma(a_{22}Q_2+a_{23}Q_3)=\sigma(Q_1)\sigma(b_{32}P_2+b_{33}P_3)}$.\\
$(ii.2)$ $\displaystyle{\sigma(a_{22}Q_2+a_{23}Q_3)\sigma(b_{32}P_2+b_{33}P_3)=\dfrac{\hbar}{2}|a_{21}b_{31}|}$.\\
$(iii\mathrm{-})$ $a_{21}\geq 0$, $b_{31}\geq 0$ and $a_{21}+b_{31}=1$.\\
From the conditions $(ii.1)$ and $(ii.2)$, we obtain
the condition $(ii)$ of the theorem.
If $a_{21}b_{31}=0$, we get $\sigma(a_{22}Q_2+a_{23}Q_3)=\sigma(b_{32}P_2+b_{33}P_3)=0$.
Since at least one of $a_{22}$, $a_{23}$, $b_{32}$ and $b_{33}$ is non-zero,
$\sigma(a_{22}Q_2+a_{23}Q_3)=\sigma(b_{32}P_2+b_{33}P_3)=0$ never holds for 
any unit vector $\xi$ of $L^2(\mathbb{R}^2)$.
Therefore, $a_{21}b_{31}\neq 0$ must be satisfied, so that we have the condition $(iii)$
of the theorem. We then have
\begin{align}
\varepsilon(Q_1)^2 &=(a_{21}-1)^2\sigma_1^2+|a_{21}b_{31}|\sigma_1^2=(1-a_{21})\sigma_1^2,\\
\eta(P_1)^2 &= (b_{31}-1)^2\hat{\sigma}_1^2+|a_{21}b_{31}|\hat{\sigma}_1^2=a_{21}\hat{\sigma}_1^2.
\end{align}

To complete the proof, for every $\nu\in(0,1)$,
we find $S$ and $\tau>0$ such that $a_{21}=\nu$ and $b_{31}=1-\nu$.
$S$ satisties $S^3=(-E)S$, so that we have
\begin{equation}
e^{tS}=\left\{
\begin{array}{ll}
\displaystyle{I+\dfrac{\sin(t\sqrt{E})}{\sqrt{E}}S
+\dfrac{1-\cos(t\sqrt{E})}{E}S^2}, & (E>0) \\
\displaystyle{I+tS+\dfrac{1}{2}t^2 S^2}, & (E=0) \\
\displaystyle{I+\dfrac{\sinh(t\sqrt{-E})}{\sqrt{-E}}S} & \\
\hspace{14mm}\displaystyle{+\dfrac{\cosh(t\sqrt{-E})-1}{-E}S^2} & (E<0)
\end{array}
\right. \label{ExSol}
\end{equation}
for all $t\in\mathbb{R}$. Independent of the sign of $E$,
$e^{\tau S}=(a_{ij})$ and $e^{-\tau S^T}=(b_{ij})$
satisfy $a_{22}=b_{33}$ and $a_{23}=b_{32}$. Since $[Q_2(\tau),P_3(\tau)]=0$,
we have $a_{21}b_{31}+2a_{22}a_{23}=0$.
Then, we use $\xi_{a_{21},a_{22}}$ as the state of $\mathbf{P}$,
i.e., $\xi_{\nu,\kappa}$ with $\nu=a_{21}$ and $\kappa=a_{22}$.
$\xi_{a_{21},a_{22}}$ is the product of two Gaussian states $\xi_2$ and $\xi_3$:
It has the form $\xi_{a_{21},a_{22}}(x_2,x_3)=\xi_2(x_2)\xi_3(x_3)$ in the coordinate representation,
where $\xi_2$ and $\xi_3$ are given by
\begin{align}
\xi_2(x_2) &=\sqrt[4]{\frac{|a_{22}|}{(2\pi)|a_{23}|\sigma_1^2}}
e^{-\frac{|a_{22}|}{4|a_{23}|\sigma_1^2}\left(x_2-\frac{1-a_{21}}{a_{22}}q_1\right)^2},\\
\xi_3(x_3) &=\sqrt[4]{\frac{|a_{23}|}{(2\pi)|a_{22}|\sigma_1^2}}
e^{-\frac{|a_{23}|}{4|a_{22}|\sigma_1^2} x_3^2+i\frac{a_{21}p_1}{a_{22} \hbar}x_3},
\end{align}
respectively, in the coordinate representation.
By Eq~(\ref{ExSol}), the cases $E>0$, $E=0$ and $E<0$ must be handled separately.

[$E>0$] Both $a_{21}=\nu$ and $b_{31}=1-\nu$ are satisfied if and only if it holds that
\begin{equation}
\dfrac{\sin(\tau\sqrt{E})}{\sqrt{E}}+
\gamma_2\dfrac{1-\cos(\tau\sqrt{E})}{E} = \dfrac{\nu}{\alpha_1}=\dfrac{1-\nu}{-\alpha_3}. \label{E1}
\end{equation}
For example, for every $0<\nu<1$, $E>0$, $\gamma_2>0$ and $0<\tau<\dfrac{\pi}{\sqrt{E}}$,
there uniquely exist $\alpha_1>0$ and $\alpha_3<0$ satisfying Eq~(\ref{E1}),
\textit{which completes the proof of the theorem}.
The family $\{\mathbb{X}_\nu\}_{\nu\in(0,1)}$ of
linear simultaneous $(Q_1,P_1)$-measurements are contained in this case.

[$E=0$] Both $a_{21}=\nu$ and $b_{31}=1-\nu$ are satisfied if and only if it holds that
\begin{equation}
\tau+\dfrac{1}{2}\gamma_2\tau^2= \dfrac{\nu}{\alpha_1}=\dfrac{1-\nu}{-\alpha_3}. \label{E2}
\end{equation}
For every $0<\nu<1$, $\gamma_2\geq 0$ and $\tau>0$,
there uniquely exist $\alpha_1>0$ and $\alpha_3<0$ satisfying Eq~(\ref{E2}).
The families $\{\mathbb{Y}_\nu^2\}_{\nu\in(0,1)}$ and $\{\mathbb{Y}_\nu^0\}_{\nu\in(0,1)}$ of
linear simultaneous $(Q_1,P_1)$-measurements are contained in this case.

[$E<0$] Both $a_{21}=\nu$ and $b_{31}=1-\nu$ are satisfied if and only if it holds that
\begin{equation}
\dfrac{\sinh(\tau\sqrt{-E})}{\sqrt{-E}}+
\gamma_2 \dfrac{\cosh(\tau\sqrt{-E})-1}{-E}= \dfrac{\nu}{\alpha_1}=\dfrac{1-\nu}{-\alpha_3}. \label{E3}
\end{equation}
For every $0<\nu<1$, $E<0$, $\gamma_2>0$ and $\tau>0$,
there uniquely exist $\alpha_1>0$ and $\alpha_3<0$ satisfying Eq~(\ref{E3}).
The family $\{\mathbb{Z}_\nu\}_{\nu\in(0,1)}$ of
linear simultaneous $(Q_1,P_1)$-measurements are contained in this case.

$e^{\tau S}=A=(a_{ij})$ and $e^{-\tau S^T}=B=(b_{ij})$ in each model are then given as follows:
\begin{table}[H]
\begin{center}
\caption{}
\begin{ruledtabular}
\begin{tabular}{ccc} 
 & $e^{\tau S}$ & $e^{-\tau S^T}$\\ \hline
$\mathbb{X}_\nu$ & $\left(
\begin{array}{ccc}
-1 &-\frac{4}{\nu} & 0 \\
\nu & 2& -\frac{\nu(1-\nu)}{4} \\
0 &-\frac{4}{\nu(1-\nu)} & 0
\end{array}
\right)$ & $\left(
\begin{array}{ccc}
-1 &0 & 0 \\
0 & 0 & -\frac{4}{\nu(1-\nu)} \\
1-\nu &-\frac{\nu(1-\nu)}{4} & 2
\end{array}
\right)$ \\ \hline
$\mathbb{Y}_\nu^2$ & $\left(
\begin{array}{ccc}
-1 & -\frac{8}{\nu}& 0 \\
\nu & 4& -\frac{\nu(1-\nu)}{8} \\
0 &-\frac{8}{\nu(1-\nu)} & 0
\end{array}
\right)$ & $\left(
\begin{array}{ccc}
-1 & 0& -\frac{8}{1-\nu} \\
0 & 0& -\frac{8}{\nu(1-\nu)} \\
1-\nu &-\frac{\nu(1-\nu)}{8} & 4
\end{array}
\right)$ \\ \hline
$\mathbb{Y}_\nu^0$ & $\left(
\begin{array}{ccc}
1 & 0& -(1-\nu) \\
\nu & 1& -\frac{\nu(1-\nu)}{2} \\
0 &0 & 1
\end{array}
\right)$ & $\left(
\begin{array}{ccc}
1 & -\nu& 0 \\
0 & 1& 0 \\
1-\nu &-\frac{\nu(1-\nu)}{2} & 1
\end{array}
\right)$ \\ \hline
$\mathbb{Z}_\nu$ & $\left(
\begin{array}{ccc}
1 & 0 & -\frac{1-\nu}{2} \\
\nu & 2& -\frac{\nu(1-\nu)}{4} \\
0 & 0& \frac{1}{2}
\end{array}
\right)$ & $\left(
\begin{array}{ccc}
1 &-\frac{\nu}{2} & 0 \\
0 & \frac{1}{2}& 0 \\
1-\nu & -\frac{\nu(1-\nu)}{4}& 2
\end{array}
\right)$ \\ 
\end{tabular}
\end{ruledtabular}
\end{center}
\end{table}

\subsection{Probability distributions and families of posterior states\label{5.2}}

The characteristic function $\lambda$ of the probability measure $\mu$ on $\mathbb{R}^d$ is defined as
the inverse Fourier transform of $\mu$:
\begin{equation}
\lambda(k)=\int_{\mathbb{R}^d} e^{i\langle x, k\rangle}\;d\mu(x),
\end{equation}
where $\langle \cdot,\cdot \rangle$ is the inner product of $\mathbb{R}^d$.
For any observables $X_1$, $X_2$ and vector state $\phi$,
the characteristic function of $\mu^{X_1,X_2}_\phi$ is denoted by $\lambda^{X_1,X_2}_\phi$.
The characteristic function of a Gaussian measure
\begin{equation}
d\mu_{V,m}(x)=\dfrac{1}{\sqrt{(2\pi)^d\det(V)}}e^{-\frac{1}{2}\langle x-m,V^{-1} (x-m)\rangle}\;dx \label{GM1}
\end{equation}
has the following form:
\begin{equation}
\lambda_{V,m}(k)=e^{i\langle m,k \rangle-\frac{1}{2} \langle k,Vk \rangle}, \label{GM2}
\end{equation}
where $V>0$ is a covariance matrix and $m\in\mathbb{R}^d$ is a mean vector.
Conversely, if a characteristic function is given by Eq.~(\ref{GM2}),
then the corresponding probability measure is a Gaussian measure given by Eq.~(\ref{GM1}).
We refer the reader to textbooks of probability theory and statistics.

The characteristic function $\lambda^{Q_1(0),Q_2(\tau)}_{\psi\otimes\xi_{a_{21},a_{22}}}$
 of $\mu^{Q_1(0),Q_2(\tau)}_{\psi\otimes\xi_{a_{21},a_{22}}}$ is given by
\begin{align}
&\hspace{5mm}\lambda^{Q_1(0),Q_2(\tau)}_{\psi\otimes\xi_{a_{21},a_{22}}}(k) \nonumber \\
&=\langle e^{ik_1Q_1(0)+ik_2Q_2(\tau)}\rangle_{\psi\otimes\xi_{a_{21},a_{22}}} \nonumber\\
 &=\langle 
  e^{i(k_1+a_{21}k_2)Q_1(0)+ia_{22}k_2Q_2(0)+ia_{23}k_2Q_3(0)}
  \rangle_{\psi\otimes\xi_{a_{21},a_{22}}} \nonumber \\
 &= \langle \psi| e^{i(k_1+a_{21}k_2)Q_1}\psi\rangle
 \langle \xi_2| e^{i(a_{22}k_2)Q_2}\xi_2\rangle\langle \xi_3| e^{i(a_{23}k_2)Q_3}\xi_3\rangle\nonumber \\
 &=e^{iq_1(k_1+a_{21}k_2)-\frac{1}{2}\sigma_1^2(k_1+a_{21}k_2)^2} \nonumber\\
&\hspace{5mm}\times
e^{i\frac{1-a_{21}}{a_{22}}q_1(a_{22}k_2)-\frac{1}{2}\left|\frac{a_{23}}{a_{22}}\right|\sigma_1^2(a_{22}k_2)^2} 
e^{-\frac{1}{2}\left|\frac{a_{22}}{a_{23}}\right|\sigma_1^2(a_{23}k_2)^2}\nonumber \\
 &= e^{iq_1k_1+iq_1k_2-
 \frac{1}{2}\sigma_1^2\left\{(k_1+a_{21}k_2)^2+ 2|a_{22}a_{23}|k_2^2\right\}} \nonumber \\
 &= \lambda_{W,q}(k)
\end{align}
for all $k= \left(
\begin{array}{c}
k_1  \\
k_2
\end{array}
\right)\in\mathbb{R}^2$,
where $q=
\left(
\begin{array}{c}
q_1  \\
q_1
\end{array}
\right)$ and $W=\sigma_1^2\left(
\begin{array}{cc}
1 & a_{21} \\
a_{21} & a_{21}
\end{array}
\right)$. Here we used $0<a_{21}<1$, $a_{21}(1-a_{21})+2a_{22}a_{23}=0$, which is obtained from
the condition $(iii)$ of the theorem and $a_{21}b_{31}+2a_{22}a_{23}=0$, and the relation
\begin{equation}
\langle \psi| e^{i(aQ_1+bP_1)}\psi\rangle
=e^{iq_1a-\frac{1}{2}\sigma_1^2a^2} e^{ip_1b-\frac{1}{2}\hat{\sigma}_1^2 b^2}
\end{equation}
for all $a,b\in\mathbb{R}$.
From $\det(W) = \sigma_1^4a_{21}(1-a_{21})$ and
\begin{equation}
W^{-1} = \dfrac{1}{(1-a_{21})\sigma_1^2}\left(
\begin{array}{cc}
1 & -1 \\
-1 & 1
\end{array}
\right)+\dfrac{1}{a_{21}\sigma_1^2}\left(
\begin{array}{cc}
0 & 0 \\
0 & 1
\end{array}
\right),
\end{equation}
we obtain 
\begin{equation}
p^{Q_1(0),Q_2(\tau)}_{\psi\otimes\xi_{a_{21},a_{22}}}(x,z)=
p_{(1-a_{21})\sigma_1^2}(x-z)p_{a_{21}\sigma_1^2}(z-q_1). \label{JPDQQX}
\end{equation}
Eq.~(\ref{JDQ1}) is obtained from Eq.~(\ref{JPDQQX}) for 
$\mathbb{X}_\nu$, $\mathbb{Y}_\nu^2$, $\mathbb{Y}_\nu^0$ and $\mathbb{Z}_\nu$. Similarly, we have
\begin{align}
p^{P_1(0),P_3(\tau)}_{\psi\otimes\xi_{a_{21},a_{22}}}(y,w) &= 
p_{a_{21}\hat{\sigma}_1^2}(y-w)p_{(1-a_{21})\hat{\sigma}_1^2}(w-p_1),  \\ 
p^{Q_2(\tau),P_3(\tau)}_{\psi\otimes\xi_{a_{21},a_{22}}}(z,w)
&= p_{a_{21}\sigma_1^2}(z-q_1)p_{(1-a_{21})\hat{\sigma}_1^2}(w-p_1). 
\end{align}
In particular, Eqs.~(\ref{JDQ2}) and (\ref{JDQ3}) are derived in the same way.

Next, for every $\nu\in(0,1)$, we find the family of posterior states for $(\mathbb{Y}_\nu^0,\psi)$.
We check the following probability density functions via their characteristic functions:
\begin{align}
&\hspace{5mm} p^{Q_1(\tau),Q_2(\tau),P_3(\tau)}_{\psi\otimes\xi_{\nu,1}}(x,z,w) \nonumber\\
&= p_{\frac{(1-\nu)\sigma_1^2}{\nu}}\left(x-\frac{z-(1-\nu)q_1}{\nu}\right) \nonumber\\
&\hspace{20mm}\times p_{\nu\sigma_1^2}(z-q_1)p_{(1-\nu)\hat{\sigma}_1^2}(w-p_1), \label{JPDAA} \\
&\hspace{5mm}p^{P_1(\tau),Q_2(\tau),P_3(\tau)}_{\psi\otimes\xi_{\nu,1}}(y,z,w) \nonumber\\
&= p_{\frac{\nu\hat{\sigma}_1^2}{1-\nu}}\left(y-\frac{w-\nu p_1}{1-\nu}\right)
p_{\nu\sigma_1^2}(z-q_1)p_{(1-\nu)\hat{\sigma}_1^2}(w-p_1). \label{JPDAB}
\end{align}
For example, the characteristic function $\lambda^{Q_1(\tau),Q_2(\tau),P_3(\tau)}_{\psi\otimes\xi_{\nu,1}}$
 of $\mu^{Q_1(\tau),Q_2(\tau),P_3(\tau)}_{\psi\otimes\xi_{\nu,1}}$ is given by
\begin{align}
&\hspace{5mm}\lambda^{Q_1(\tau),Q_2(\tau),P_3(\tau)}_{\psi\otimes\xi_{\nu,1}}(k) \nonumber \\
&=\langle e^{ik_1Q_1(\tau)+ik_2Q_2(\tau)+ik_3P_3(\tau)}\rangle_{\psi\otimes\xi_{\nu,1}} \nonumber\\
 &= \langle \psi| e^{i(k_1+\nu k_2)Q_1+i(1-\nu)k_3 P_1}\psi\rangle
 \langle \xi_2| e^{ik_2 Q_2-i \frac{\nu(1-\nu)}{2}k_3 P_2}\xi_2\rangle \nonumber \\
 &\hspace{5mm}\times \langle \xi_3| e^{-i(1-\nu)k_1Q_3-i\frac{\nu(1-\nu)}{2}k_2Q_3+i k_3P_3}\xi_3\rangle\nonumber \\
 &=e^{iq_1(k_1+\nu k_2)-\frac{1}{2}\sigma_1^2(k_1+\nu k_2)^2}
 e^{ip_1(1-\nu)k_3-\frac{1}{2}\hat{\sigma}_1^2(1-\nu)^2 k_3^2} \nonumber\\
&\hspace{5mm}\times
e^{i(1-\nu)q_1k_2-\frac{1}{2}\frac{\nu(1-\nu)}{2}\sigma_1^2 k_2^2} 
e^{-\frac{1}{2} \frac{2}{\nu(1-\nu)}\hat{\sigma}_1^2\left(\frac{\nu(1-\nu)}{2}k_3\right)^2}\nonumber \\
&\hspace{5mm}\times
e^{-\frac{1}{2}\frac{2}{\nu(1-\nu)}\sigma_1^2\left(-(1-\nu)k_1-\frac{\nu(1-\nu)}{2}k_2\right)^2} 
e^{i\nu p_1k_3 -\frac{1}{2}\frac{\nu(1-\nu)}{2}\hat{\sigma}_1^2k_3^2}\nonumber \\
 &= \lambda_{Z,q}(\tilde{k}) \lambda_{(1-\nu)\hat{\sigma}_1^2,p_1}(k_3)
\end{align}
for all $k=\left(
\begin{array}{c}
k_1  \\
k_2  \\
k_3 
\end{array}
\right)\in\mathbb{R}^3$, where $\tilde{k}=\left(
\begin{array}{c}
k_1  \\
k_2
\end{array}
\right)$ and $Z=\sigma_1^2\left(
\begin{array}{cc}
\frac{2-\nu}{\nu} & 1 \\
1 & \nu
\end{array}
\right)$. From $\det Z=(1-\nu)\sigma_1^4$ and
\begin{equation}
Z^{-1}=\frac{\nu}{(1-\nu)\sigma_1^2}\left(
\begin{array}{cc}
1 & -\frac{1}{\nu} \\
-\frac{1}{\nu} & \frac{1}{\nu^2}
\end{array}
\right)+\frac{1}{\nu\sigma_1^2}\left(
\begin{array}{cc}
0 & 0 \\
0 & 1
\end{array}
\right),
\end{equation}
we obtain Eq.~(\ref{JPDAA}).
The relation $\dfrac{(1-\nu)\sigma_1^2}{\nu}\cdot \dfrac{\nu\hat{\sigma}_1^2}{1-\nu}=\dfrac{\hbar^2}{4}$
implies that the family of posterior states for $(\mathbb{Y}_\nu^0,\psi)$ is given by Eq.~(\ref{FPSY})
 and is unique up to phase. For every $\nu\in(0,1)$,
the family of posterior states for $(\mathbb{Z}_\nu,\psi)$ is derived in the same way.

For every rectangular(, more generally, Borel subset) $J$ in $\mathbb{R}^2$,
we then obtain the state $\rho_J$ after the measurement under the condition that
output values not contained in $J$ is excluded, which is given by
\begin{equation}
\mathrm{Tr}[X\rho_J]=\dfrac{\langle U(\tau)(\psi\otimes\xi)|X E(J)U(\tau)(\psi\otimes\xi)\rangle}{
\langle U(\tau)(\psi\otimes\xi)| E(J)U(\tau)(\psi\otimes\xi) \rangle}
\end{equation}
whenever $\langle U(\psi\otimes\xi)|(1\otimes E(J))U(\psi\otimes\xi) \rangle\neq 0$.
Here $E$ is the spectral measure of $\mathbb{R}^2$ on $L^2(\mathbb{R}^2)$
such that $E(J_1\times J_2)=E^{Q_2}(J_1)E^{P_3}(J_2)$
for all Borel sets $J_1,J_2$ of $\mathbb{R}$.
For every $\nu\in(0,1)$,
the family $\{\psi_y\}_{y\in\mathbb{R}^2}$ of posterior states for $(\mathbb{Y}^0_\nu,\psi)$ satisfies
\begin{equation}
\rho_{J}=\dfrac{1}{\mu_{V_\nu,r}(J)}\int_J |\psi_y\rangle\langle\psi_y|\;
d\mu_{V_\nu,r}(y)
\end{equation}
for all Borel set $J$ of $\mathbb{R}^2$, where 
$V_\nu=\left(
\begin{array}{cc}
\nu\sigma_1^2 & 0 \\
0 & (1-\nu)\hat{\sigma}_1^2
\end{array}
\right)$ and $r=\left(
\begin{array}{c}
q_1  \\
p_1
\end{array}
\right)$.

\section{Summary and Perspectives\label{6}}

We have given a necessary and sufficient condition for a linear simultaneous $(Q_1,P_1)$-measurement
to satisfy Eq.~(\ref{BOQP2}), and constructed four families
$\{\mathbb{X}_\nu\}_{(0,1)}$, $\{\mathbb{Y}_\nu^2\}_{(0,1)}$, $\{\mathbb{Y}_\nu^0\}_{(0,1)}$ and $\{\mathbb{Z}_\nu\}_{(0,1)}$
of linear simultaneous $(Q_1,P_1)$-measurements.
Furthermore, we have probability distributions when using
$\{\mathbb{X}_\nu\}_{(0,1)}$, $\{\mathbb{Y}_\nu^2\}_{(0,1)}$, $\{\mathbb{Y}_\nu^0\}_{(0,1)}$ and $\{\mathbb{Z}_\nu\}_{(0,1)}$,
and families of posterior states for $\{\mathbb{Y}_\nu^0\}_{(0,1)}$ and $\{\mathbb{Z}_\nu\}_{(0,1)}$.
We believe that the results of the paper
have important implications for future research on simultaneous measurements.
There are not so many studies on simultaneous measurements
of position and momentum since Heisenberg's paper in spite of their importance.
In fact, this paper shows that there is still room for
studying simultaneous measurements of position and momentum.
The same is true for simultaneous measurements of different components of the spin.
It is desirable to study simultaneous measurements more and more actively,
in connection with the recent progress of uncertainty relations.
We believe that it will make a significant contribution to the resolution of various problems
in the field of quantum information through quantitative analysis.
In the future, the theory of simultaneous measurements of imcompatible observables
will be widely applied to quantum information processing.

\begin{acknowledgments}
The author thanks Prof. Motoichi Ohtsu and Prof. Fumio Hiroshima for their warmful encouragements.
\end{acknowledgments}



\bibliography{sample}

\begin{thebibliography}{42}%
\makeatletter
\providecommand \@ifxundefined [1]{%
 \@ifx{#1\undefined}
}%
\providecommand \@ifnum [1]{%
 \ifnum #1\expandafter \@firstoftwo
 \else \expandafter \@secondoftwo
 \fi
}%
\providecommand \@ifx [1]{%
 \ifx #1\expandafter \@firstoftwo
 \else \expandafter \@secondoftwo
 \fi
}%
\providecommand \natexlab [1]{#1}%
\providecommand \enquote  [1]{``#1''}%
\providecommand \bibnamefont  [1]{#1}%
\providecommand \bibfnamefont [1]{#1}%
\providecommand \citenamefont [1]{#1}%
\providecommand \href@noop [0]{\@secondoftwo}%
\providecommand \href [0]{\begingroup \@sanitize@url \@href}%
\providecommand \@href[1]{\@@startlink{#1}\@@href}%
\providecommand \@@href[1]{\endgroup#1\@@endlink}%
\providecommand \@sanitize@url [0]{\catcode `\\12\catcode `\$12\catcode
  `\&12\catcode `\#12\catcode `\^12\catcode `\_12\catcode `\%12\relax}%
\providecommand \@@startlink[1]{}%
\providecommand \@@endlink[0]{}%
\providecommand \url  [0]{\begingroup\@sanitize@url \@url }%
\providecommand \@url [1]{\endgroup\@href {#1}{\urlprefix }}%
\providecommand \urlprefix  [0]{URL }%
\providecommand \Eprint [0]{\href }%
\providecommand \doibase [0]{https://doi.org/}%
\providecommand \selectlanguage [0]{\@gobble}%
\providecommand \bibinfo  [0]{\@secondoftwo}%
\providecommand \bibfield  [0]{\@secondoftwo}%
\providecommand \translation [1]{[#1]}%
\providecommand \BibitemOpen [0]{}%
\providecommand \bibitemStop [0]{}%
\providecommand \bibitemNoStop [0]{.\EOS\space}%
\providecommand \EOS [0]{\spacefactor3000\relax}%
\providecommand \BibitemShut  [1]{\csname bibitem#1\endcsname}%
\let\auto@bib@innerbib\@empty
\bibitem [{\citenamefont {Heisenberg}(1927)}]{Heisenberg1927}%
  \BibitemOpen
  \bibfield  {author} {\bibinfo {author} {\bibfnamefont {W.}~\bibnamefont
  {Heisenberg}},\ }\bibfield  {title} {\bibinfo {title} {\"{U}ber den
  anschaulichen inhalt der quantentheoretischen kinematik und mechanik},\
  }\href {https://doi.org/10.1007/BF01397280} {\bibfield  {journal} {\bibinfo
  {journal} {Z. Phys.}\ }\textbf {\bibinfo {volume} {43}},\ \bibinfo {pages}
  {172} (\bibinfo {year} {1927})}\BibitemShut {NoStop}%
\bibitem [{\citenamefont {von Neumann}(2018)}]{von2018mathematical}%
  \BibitemOpen
  \bibfield  {author} {\bibinfo {author} {\bibfnamefont {J.}~\bibnamefont {von
  Neumann}},\ }\href@noop {} {\emph {\bibinfo {title} {Mathematical foundations
  of quantum mechanics: New edition}}}\ (\bibinfo  {publisher} {Princeton UP},\
  \bibinfo {address} {Princeton},\ \bibinfo {year} {2018})\BibitemShut
  {NoStop}%
\bibitem [{\citenamefont {Ozawa}(2003{\natexlab{a}})}]{ozawa2003universally}%
  \BibitemOpen
  \bibfield  {author} {\bibinfo {author} {\bibfnamefont {M.}~\bibnamefont
  {Ozawa}},\ }\bibfield  {title} {\bibinfo {title} {Universally valid
  reformulation of the \text{H}eisenberg uncertainty principle on noise and
  disturbance in measurement},\ }\href@noop {} {\bibfield  {journal} {\bibinfo
  {journal} {Phys. Rev. A}\ }\textbf {\bibinfo {volume} {67}},\ \bibinfo
  {pages} {042105} (\bibinfo {year} {2003}{\natexlab{a}})}\BibitemShut
  {NoStop}%
\bibitem [{\citenamefont {Ozawa}(2003{\natexlab{b}})}]{ozawa2003physical}%
  \BibitemOpen
  \bibfield  {author} {\bibinfo {author} {\bibfnamefont {M.}~\bibnamefont
  {Ozawa}},\ }\bibfield  {title} {\bibinfo {title} {Physical content of
  \text{H}eisenberg's uncertainty relation: limitation and reformulation},\
  }\href@noop {} {\bibfield  {journal} {\bibinfo  {journal} {Phys. Lett. A}\
  }\textbf {\bibinfo {volume} {318}},\ \bibinfo {pages} {21} (\bibinfo {year}
  {2003}{\natexlab{b}})}\BibitemShut {NoStop}%
\bibitem [{\citenamefont {Ozawa}(2004{\natexlab{a}})}]{ozawa2004uncertainty2}%
  \BibitemOpen
  \bibfield  {author} {\bibinfo {author} {\bibfnamefont {M.}~\bibnamefont
  {Ozawa}},\ }\bibfield  {title} {\bibinfo {title} {Uncertainty relations for
  joint measurements of noncommuting observables},\ }\href@noop {} {\bibfield
  {journal} {\bibinfo  {journal} {Phys. Lett. A}\ }\textbf {\bibinfo {volume}
  {320}},\ \bibinfo {pages} {367} (\bibinfo {year}
  {2004}{\natexlab{a}})}\BibitemShut {NoStop}%
\bibitem [{\citenamefont {Ozawa}(2004{\natexlab{b}})}]{ozawa2004uncertainty}%
  \BibitemOpen
  \bibfield  {author} {\bibinfo {author} {\bibfnamefont {M.}~\bibnamefont
  {Ozawa}},\ }\bibfield  {title} {\bibinfo {title} {Uncertainty relations for
  noise and disturbance in generalized quantum measurements},\ }\href@noop {}
  {\bibfield  {journal} {\bibinfo  {journal} {Ann. Phys. (N.Y.)}\ }\textbf
  {\bibinfo {volume} {311}},\ \bibinfo {pages} {350} (\bibinfo {year}
  {2004}{\natexlab{b}})}\BibitemShut {NoStop}%
\bibitem [{\citenamefont {Hall}(2004)}]{hall2004prior}%
  \BibitemOpen
  \bibfield  {author} {\bibinfo {author} {\bibfnamefont {M.~J.}\ \bibnamefont
  {Hall}},\ }\bibfield  {title} {\bibinfo {title} {Prior information: How to
  circumvent the standard joint-measurement uncertainty relation},\ }\href@noop
  {} {\bibfield  {journal} {\bibinfo  {journal} {Phys. Rev. A}\ }\textbf
  {\bibinfo {volume} {69}},\ \bibinfo {pages} {052113} (\bibinfo {year}
  {2004})}\BibitemShut {NoStop}%
\bibitem [{\citenamefont {Busch}\ \emph {et~al.}(2007)\citenamefont {Busch},
  \citenamefont {Heinonen},\ and\ \citenamefont {Lahti}}]{busch2007heisenberg}%
  \BibitemOpen
  \bibfield  {author} {\bibinfo {author} {\bibfnamefont {P.}~\bibnamefont
  {Busch}}, \bibinfo {author} {\bibfnamefont {T.}~\bibnamefont {Heinonen}},\
  and\ \bibinfo {author} {\bibfnamefont {P.}~\bibnamefont {Lahti}},\ }\bibfield
   {title} {\bibinfo {title} {Heisenberg's uncertainty principle},\ }\href@noop
  {} {\bibfield  {journal} {\bibinfo  {journal} {Phys. Rep.}\ }\textbf
  {\bibinfo {volume} {452}},\ \bibinfo {pages} {155} (\bibinfo {year}
  {2007})}\BibitemShut {NoStop}%
\bibitem [{\citenamefont {Lund}\ and\ \citenamefont
  {Wiseman}(2010)}]{lund2010measuring}%
  \BibitemOpen
  \bibfield  {author} {\bibinfo {author} {\bibfnamefont {A.}~\bibnamefont
  {Lund}}\ and\ \bibinfo {author} {\bibfnamefont {H.}~\bibnamefont {Wiseman}},\
  }\bibfield  {title} {\bibinfo {title} {Measuring measurement-disturbance
  relationships with weak values},\ }\href@noop {} {\bibfield  {journal}
  {\bibinfo  {journal} {New J. Phys.}\ }\textbf {\bibinfo {volume} {12}},\
  \bibinfo {pages} {093011} (\bibinfo {year} {2010})}\BibitemShut {NoStop}%
\bibitem [{\citenamefont {Branciard}(2013)}]{branciard2013error}%
  \BibitemOpen
  \bibfield  {author} {\bibinfo {author} {\bibfnamefont {C.}~\bibnamefont
  {Branciard}},\ }\bibfield  {title} {\bibinfo {title} {Error-tradeoff and
  error-disturbance relations for incompatible quantum measurements},\
  }\href@noop {} {\bibfield  {journal} {\bibinfo  {journal} {Proc. Nat. Acad.
  Sci.}\ }\textbf {\bibinfo {volume} {110}},\ \bibinfo {pages} {6742} (\bibinfo
  {year} {2013})}\BibitemShut {NoStop}%
\bibitem [{\citenamefont {Branciard}(2014)}]{branciard2014deriving}%
  \BibitemOpen
  \bibfield  {author} {\bibinfo {author} {\bibfnamefont {C.}~\bibnamefont
  {Branciard}},\ }\bibfield  {title} {\bibinfo {title} {Deriving tight
  error-trade-off relations for approximate joint measurements of incompatible
  quantum observables},\ }\href@noop {} {\bibfield  {journal} {\bibinfo
  {journal} {Phys. Rev. A}\ }\textbf {\bibinfo {volume} {89}},\ \bibinfo
  {pages} {022124} (\bibinfo {year} {2014})}\BibitemShut {NoStop}%
\bibitem [{\citenamefont {Ozawa}(2014)}]{ozawa2014errordisturbance2}%
  \BibitemOpen
  \bibfield  {author} {\bibinfo {author} {\bibfnamefont {M.}~\bibnamefont
  {Ozawa}},\ }\bibfield  {title} {\bibinfo {title} {Error-disturbance relations
  in mixed states}\ }(\bibinfo {year} {2014})\ \Eprint
  {https://arxiv.org/abs/Preprint at https://arxiv.org/abs/1404.3388} {Preprint
  at https://arxiv.org/abs/1404.3388} \BibitemShut {NoStop}%
\bibitem [{\citenamefont {Busch}\ \emph {et~al.}(2013)\citenamefont {Busch},
  \citenamefont {Lahti},\ and\ \citenamefont {Werner}}]{busch2013proof}%
  \BibitemOpen
  \bibfield  {author} {\bibinfo {author} {\bibfnamefont {P.}~\bibnamefont
  {Busch}}, \bibinfo {author} {\bibfnamefont {P.}~\bibnamefont {Lahti}},\ and\
  \bibinfo {author} {\bibfnamefont {R.~F.}\ \bibnamefont {Werner}},\ }\bibfield
   {title} {\bibinfo {title} {Proof of \text{H}eisenberg’s error-disturbance
  relation},\ }\href@noop {} {\bibfield  {journal} {\bibinfo  {journal} {Phys.
  Rev. Lett.}\ }\textbf {\bibinfo {volume} {111}},\ \bibinfo {pages} {160405}
  (\bibinfo {year} {2013})}\BibitemShut {NoStop}%
\bibitem [{\citenamefont {Busch}\ \emph
  {et~al.}(2014{\natexlab{a}})\citenamefont {Busch}, \citenamefont {Lahti},\
  and\ \citenamefont {Werner}}]{busch2014colloquium}%
  \BibitemOpen
  \bibfield  {author} {\bibinfo {author} {\bibfnamefont {P.}~\bibnamefont
  {Busch}}, \bibinfo {author} {\bibfnamefont {P.}~\bibnamefont {Lahti}},\ and\
  \bibinfo {author} {\bibfnamefont {R.~F.}\ \bibnamefont {Werner}},\ }\bibfield
   {title} {\bibinfo {title} {\textit{Colloquium}: Quantum root-mean-square
  error and measurement uncertainty relations},\ }\href@noop {} {\bibfield
  {journal} {\bibinfo  {journal} {Rev. Mod. Phys.}\ }\textbf {\bibinfo {volume}
  {86}},\ \bibinfo {pages} {1261} (\bibinfo {year}
  {2014}{\natexlab{a}})}\BibitemShut {NoStop}%
\bibitem [{\citenamefont {Busch}\ \emph
  {et~al.}(2014{\natexlab{b}})\citenamefont {Busch}, \citenamefont {Lahti},\
  and\ \citenamefont {Werner}}]{busch2014measurement}%
  \BibitemOpen
  \bibfield  {author} {\bibinfo {author} {\bibfnamefont {P.}~\bibnamefont
  {Busch}}, \bibinfo {author} {\bibfnamefont {P.}~\bibnamefont {Lahti}},\ and\
  \bibinfo {author} {\bibfnamefont {R.~F.}\ \bibnamefont {Werner}},\ }\bibfield
   {title} {\bibinfo {title} {Measurement uncertainty relations},\ }\href@noop
  {} {\bibfield  {journal} {\bibinfo  {journal} {J. Math. Phys.}\ }\textbf
  {\bibinfo {volume} {55}},\ \bibinfo {pages} {042111} (\bibinfo {year}
  {2014}{\natexlab{b}})}\BibitemShut {NoStop}%
\bibitem [{\citenamefont {Ozawa}(2013)}]{ozawa2013disproving}%
  \BibitemOpen
  \bibfield  {author} {\bibinfo {author} {\bibfnamefont {M.}~\bibnamefont
  {Ozawa}},\ }\href@noop {} {\bibinfo {title} {Disproving \text{H}eisenberg's
  error-disturbance relation}} (\bibinfo {year} {2013}),\ \Eprint
  {https://arxiv.org/abs/1308.3540} {arXiv:1308.3540 [quant-ph]} \BibitemShut
  {NoStop}%
\bibitem [{\citenamefont {Buscemi}\ \emph {et~al.}(2014)\citenamefont
  {Buscemi}, \citenamefont {Hall}, \citenamefont {Ozawa},\ and\ \citenamefont
  {Wilde}}]{buscemi2014noise}%
  \BibitemOpen
  \bibfield  {author} {\bibinfo {author} {\bibfnamefont {F.}~\bibnamefont
  {Buscemi}}, \bibinfo {author} {\bibfnamefont {M.~J.}\ \bibnamefont {Hall}},
  \bibinfo {author} {\bibfnamefont {M.}~\bibnamefont {Ozawa}},\ and\ \bibinfo
  {author} {\bibfnamefont {M.~M.}\ \bibnamefont {Wilde}},\ }\bibfield  {title}
  {\bibinfo {title} {Noise and disturbance in quantum measurements: an
  information-theoretic approach},\ }\href@noop {} {\bibfield  {journal}
  {\bibinfo  {journal} {Phys. Rev. Lett.}\ }\textbf {\bibinfo {volume} {112}},\
  \bibinfo {pages} {050401} (\bibinfo {year} {2014})}\BibitemShut {NoStop}%
\bibitem [{\citenamefont {Dressel}\ and\ \citenamefont
  {Nori}(2014)}]{dressel2014certainty}%
  \BibitemOpen
  \bibfield  {author} {\bibinfo {author} {\bibfnamefont {J.}~\bibnamefont
  {Dressel}}\ and\ \bibinfo {author} {\bibfnamefont {F.}~\bibnamefont {Nori}},\
  }\bibfield  {title} {\bibinfo {title} {Certainty in \text{H}eisenberg's
  uncertainty principle: revisiting definitions for estimation errors and
  disturbance},\ }\href@noop {} {\bibfield  {journal} {\bibinfo  {journal}
  {Phys. Rev. A}\ }\textbf {\bibinfo {volume} {89}},\ \bibinfo {pages} {022106}
  (\bibinfo {year} {2014})}\BibitemShut {NoStop}%
\bibitem [{\citenamefont {Korzekwa}\ \emph {et~al.}(2014)\citenamefont
  {Korzekwa}, \citenamefont {Jennings},\ and\ \citenamefont
  {Rudolph}}]{korzekwa2014operational}%
  \BibitemOpen
  \bibfield  {author} {\bibinfo {author} {\bibfnamefont {K.}~\bibnamefont
  {Korzekwa}}, \bibinfo {author} {\bibfnamefont {D.}~\bibnamefont {Jennings}},\
  and\ \bibinfo {author} {\bibfnamefont {T.}~\bibnamefont {Rudolph}},\
  }\bibfield  {title} {\bibinfo {title} {Operational constraints on
  state-dependent formulations of quantum error-disturbance trade-off
  relations},\ }\href@noop {} {\bibfield  {journal} {\bibinfo  {journal} {Phys.
  Rev. A}\ }\textbf {\bibinfo {volume} {89}},\ \bibinfo {pages} {052108}
  (\bibinfo {year} {2014})}\BibitemShut {NoStop}%
\bibitem [{\citenamefont {Erhart}\ \emph {et~al.}(2012)\citenamefont {Erhart},
  \citenamefont {Sponar}, \citenamefont {Sulyok}, \citenamefont {Badurek},
  \citenamefont {Ozawa},\ and\ \citenamefont
  {Hasegawa}}]{erhart2012experimental}%
  \BibitemOpen
  \bibfield  {author} {\bibinfo {author} {\bibfnamefont {J.}~\bibnamefont
  {Erhart}}, \bibinfo {author} {\bibfnamefont {S.}~\bibnamefont {Sponar}},
  \bibinfo {author} {\bibfnamefont {G.}~\bibnamefont {Sulyok}}, \bibinfo
  {author} {\bibfnamefont {G.}~\bibnamefont {Badurek}}, \bibinfo {author}
  {\bibfnamefont {M.}~\bibnamefont {Ozawa}},\ and\ \bibinfo {author}
  {\bibfnamefont {Y.}~\bibnamefont {Hasegawa}},\ }\bibfield  {title} {\bibinfo
  {title} {Experimental demonstration of a universally valid error-disturbance
  uncertainty relation in spin measurements},\ }\href@noop {} {\bibfield
  {journal} {\bibinfo  {journal} {Nature Phys.}\ }\textbf {\bibinfo {volume}
  {8}},\ \bibinfo {pages} {185} (\bibinfo {year} {2012})}\BibitemShut {NoStop}%
\bibitem [{\citenamefont {Sulyok}\ \emph {et~al.}(2013)\citenamefont {Sulyok},
  \citenamefont {Sponar}, \citenamefont {Erhart}, \citenamefont {Badurek},
  \citenamefont {Ozawa},\ and\ \citenamefont {Hasegawa}}]{sulyok2013violation}%
  \BibitemOpen
  \bibfield  {author} {\bibinfo {author} {\bibfnamefont {G.}~\bibnamefont
  {Sulyok}}, \bibinfo {author} {\bibfnamefont {S.}~\bibnamefont {Sponar}},
  \bibinfo {author} {\bibfnamefont {J.}~\bibnamefont {Erhart}}, \bibinfo
  {author} {\bibfnamefont {G.}~\bibnamefont {Badurek}}, \bibinfo {author}
  {\bibfnamefont {M.}~\bibnamefont {Ozawa}},\ and\ \bibinfo {author}
  {\bibfnamefont {Y.}~\bibnamefont {Hasegawa}},\ }\bibfield  {title} {\bibinfo
  {title} {Violation of \text{H}eisenberg's error-disturbance uncertainty
  relation in neutron-spin measurements},\ }\href@noop {} {\bibfield  {journal}
  {\bibinfo  {journal} {Phys. Rev. A}\ }\textbf {\bibinfo {volume} {88}},\
  \bibinfo {pages} {022110} (\bibinfo {year} {2013})}\BibitemShut {NoStop}%
\bibitem [{\citenamefont {Baek}\ \emph {et~al.}(2013)\citenamefont {Baek},
  \citenamefont {Kaneda}, \citenamefont {Ozawa},\ and\ \citenamefont
  {Edamatsu}}]{baek2013experimental}%
  \BibitemOpen
  \bibfield  {author} {\bibinfo {author} {\bibfnamefont {S.-Y.}\ \bibnamefont
  {Baek}}, \bibinfo {author} {\bibfnamefont {F.}~\bibnamefont {Kaneda}},
  \bibinfo {author} {\bibfnamefont {M.}~\bibnamefont {Ozawa}},\ and\ \bibinfo
  {author} {\bibfnamefont {K.}~\bibnamefont {Edamatsu}},\ }\bibfield  {title}
  {\bibinfo {title} {Experimental violation and reformulation of the
  \text{H}eisenberg's error-disturbance uncertainty relation},\ }\href@noop {}
  {\bibfield  {journal} {\bibinfo  {journal} {Sci. Rep.}\ }\textbf {\bibinfo
  {volume} {3}},\ \bibinfo {pages} {2221} (\bibinfo {year} {2013})}\BibitemShut
  {NoStop}%
\bibitem [{\citenamefont {Kaneda}\ \emph {et~al.}(2014)\citenamefont {Kaneda},
  \citenamefont {Baek}, \citenamefont {Ozawa},\ and\ \citenamefont
  {Edamatsu}}]{kaneda2014experimental}%
  \BibitemOpen
  \bibfield  {author} {\bibinfo {author} {\bibfnamefont {F.}~\bibnamefont
  {Kaneda}}, \bibinfo {author} {\bibfnamefont {S.-Y.}\ \bibnamefont {Baek}},
  \bibinfo {author} {\bibfnamefont {M.}~\bibnamefont {Ozawa}},\ and\ \bibinfo
  {author} {\bibfnamefont {K.}~\bibnamefont {Edamatsu}},\ }\bibfield  {title}
  {\bibinfo {title} {Experimental test of error-disturbance uncertainty
  relations by weak measurement},\ }\href@noop {} {\bibfield  {journal}
  {\bibinfo  {journal} {Phys. Rev. Lett.}\ }\textbf {\bibinfo {volume} {112}},\
  \bibinfo {pages} {020402} (\bibinfo {year} {2014})}\BibitemShut {NoStop}%
\bibitem [{\citenamefont {Ringbauer}\ \emph {et~al.}(2014)\citenamefont
  {Ringbauer}, \citenamefont {Biggerstaff}, \citenamefont {Broome},
  \citenamefont {Fedrizzi}, \citenamefont {Branciard},\ and\ \citenamefont
  {White}}]{ringbauer2014experimental}%
  \BibitemOpen
  \bibfield  {author} {\bibinfo {author} {\bibfnamefont {M.}~\bibnamefont
  {Ringbauer}}, \bibinfo {author} {\bibfnamefont {D.~N.}\ \bibnamefont
  {Biggerstaff}}, \bibinfo {author} {\bibfnamefont {M.~A.}\ \bibnamefont
  {Broome}}, \bibinfo {author} {\bibfnamefont {A.}~\bibnamefont {Fedrizzi}},
  \bibinfo {author} {\bibfnamefont {C.}~\bibnamefont {Branciard}},\ and\
  \bibinfo {author} {\bibfnamefont {A.~G.}\ \bibnamefont {White}},\ }\bibfield
  {title} {\bibinfo {title} {Experimental joint quantum measurements with
  minimum uncertainty},\ }\href@noop {} {\bibfield  {journal} {\bibinfo
  {journal} {Phys. Rev. Lett.}\ }\textbf {\bibinfo {volume} {112}},\ \bibinfo
  {pages} {020401} (\bibinfo {year} {2014})}\BibitemShut {NoStop}%
\bibitem [{\citenamefont {Sulyok}\ \emph {et~al.}(2015)\citenamefont {Sulyok},
  \citenamefont {Sponar}, \citenamefont {Demirel}, \citenamefont {Buscemi},
  \citenamefont {Hall}, \citenamefont {Ozawa},\ and\ \citenamefont
  {Hasegawa}}]{sulyok2015experimental}%
  \BibitemOpen
  \bibfield  {author} {\bibinfo {author} {\bibfnamefont {G.}~\bibnamefont
  {Sulyok}}, \bibinfo {author} {\bibfnamefont {S.}~\bibnamefont {Sponar}},
  \bibinfo {author} {\bibfnamefont {B.}~\bibnamefont {Demirel}}, \bibinfo
  {author} {\bibfnamefont {F.}~\bibnamefont {Buscemi}}, \bibinfo {author}
  {\bibfnamefont {M.~J.}\ \bibnamefont {Hall}}, \bibinfo {author}
  {\bibfnamefont {M.}~\bibnamefont {Ozawa}},\ and\ \bibinfo {author}
  {\bibfnamefont {Y.}~\bibnamefont {Hasegawa}},\ }\bibfield  {title} {\bibinfo
  {title} {Experimental test of entropic noise-disturbance uncertainty
  relations for spin-1/2 measurements},\ }\href@noop {} {\bibfield  {journal}
  {\bibinfo  {journal} {Phys. Rev. Lett.}\ }\textbf {\bibinfo {volume} {115}},\
  \bibinfo {pages} {030401} (\bibinfo {year} {2015})}\BibitemShut {NoStop}%
\bibitem [{\citenamefont {Demirel}\ \emph {et~al.}(2016)\citenamefont
  {Demirel}, \citenamefont {Sponar}, \citenamefont {Sulyok}, \citenamefont
  {Ozawa},\ and\ \citenamefont {Hasegawa}}]{demirel2016experimental}%
  \BibitemOpen
  \bibfield  {author} {\bibinfo {author} {\bibfnamefont {B.}~\bibnamefont
  {Demirel}}, \bibinfo {author} {\bibfnamefont {S.}~\bibnamefont {Sponar}},
  \bibinfo {author} {\bibfnamefont {G.}~\bibnamefont {Sulyok}}, \bibinfo
  {author} {\bibfnamefont {M.}~\bibnamefont {Ozawa}},\ and\ \bibinfo {author}
  {\bibfnamefont {Y.}~\bibnamefont {Hasegawa}},\ }\bibfield  {title} {\bibinfo
  {title} {Experimental test of residual error-disturbance uncertainty
  relations for mixed spin-$1/2$ states},\ }\href@noop {} {\bibfield  {journal}
  {\bibinfo  {journal} {Phys. Rev. Lett.}\ }\textbf {\bibinfo {volume} {117}},\
  \bibinfo {pages} {140402} (\bibinfo {year} {2016})}\BibitemShut {NoStop}%
\bibitem [{\citenamefont {Demirel}\ \emph {et~al.}(2019)\citenamefont
  {Demirel}, \citenamefont {Sponar}, \citenamefont {Abbott}, \citenamefont
  {Branciard},\ and\ \citenamefont {Hasegawa}}]{demirel2019experimental}%
  \BibitemOpen
  \bibfield  {author} {\bibinfo {author} {\bibfnamefont {B.}~\bibnamefont
  {Demirel}}, \bibinfo {author} {\bibfnamefont {S.}~\bibnamefont {Sponar}},
  \bibinfo {author} {\bibfnamefont {A.~A.}\ \bibnamefont {Abbott}}, \bibinfo
  {author} {\bibfnamefont {C.}~\bibnamefont {Branciard}},\ and\ \bibinfo
  {author} {\bibfnamefont {Y.}~\bibnamefont {Hasegawa}},\ }\bibfield  {title}
  {\bibinfo {title} {Experimental test of an entropic measurement uncertainty
  relation for arbitrary qubit observables},\ }\href@noop {} {\bibfield
  {journal} {\bibinfo  {journal} {New J. Phys.}\ }\textbf {\bibinfo {volume}
  {21}},\ \bibinfo {pages} {013038} (\bibinfo {year} {2019})}\BibitemShut
  {NoStop}%
\bibitem [{\citenamefont {Liu}\ \emph {et~al.}(2019{\natexlab{a}})\citenamefont
  {Liu}, \citenamefont {Ma}, \citenamefont {Kang}, \citenamefont {Han},
  \citenamefont {Wang}, \citenamefont {Qin}, \citenamefont {Su},\ and\
  \citenamefont {Peng}}]{liu2019experimental}%
  \BibitemOpen
  \bibfield  {author} {\bibinfo {author} {\bibfnamefont {Y.}~\bibnamefont
  {Liu}}, \bibinfo {author} {\bibfnamefont {Z.}~\bibnamefont {Ma}}, \bibinfo
  {author} {\bibfnamefont {H.}~\bibnamefont {Kang}}, \bibinfo {author}
  {\bibfnamefont {D.}~\bibnamefont {Han}}, \bibinfo {author} {\bibfnamefont
  {M.}~\bibnamefont {Wang}}, \bibinfo {author} {\bibfnamefont {Z.}~\bibnamefont
  {Qin}}, \bibinfo {author} {\bibfnamefont {X.}~\bibnamefont {Su}},\ and\
  \bibinfo {author} {\bibfnamefont {K.}~\bibnamefont {Peng}},\ }\bibfield
  {title} {\bibinfo {title} {Experimental test of error-tradeoff uncertainty
  relation using a continuous-variable entangled state},\ }\href@noop {}
  {\bibfield  {journal} {\bibinfo  {journal} {npj Quantum Inf.}\ }\textbf
  {\bibinfo {volume} {5}},\ \bibinfo {pages} {68} (\bibinfo {year}
  {2019}{\natexlab{a}})}\BibitemShut {NoStop}%
\bibitem [{\citenamefont {Liu}\ \emph {et~al.}(2019{\natexlab{b}})\citenamefont
  {Liu}, \citenamefont {Kang}, \citenamefont {Han}, \citenamefont {Su},\ and\
  \citenamefont {Peng}}]{liu2019experimental2}%
  \BibitemOpen
  \bibfield  {author} {\bibinfo {author} {\bibfnamefont {Y.}~\bibnamefont
  {Liu}}, \bibinfo {author} {\bibfnamefont {H.}~\bibnamefont {Kang}}, \bibinfo
  {author} {\bibfnamefont {D.}~\bibnamefont {Han}}, \bibinfo {author}
  {\bibfnamefont {X.}~\bibnamefont {Su}},\ and\ \bibinfo {author}
  {\bibfnamefont {K.}~\bibnamefont {Peng}},\ }\bibfield  {title} {\bibinfo
  {title} {Experimental test of error-disturbance uncertainty relation with
  continuous variables},\ }\href@noop {} {\bibfield  {journal} {\bibinfo
  {journal} {Photonics Res.}\ }\textbf {\bibinfo {volume} {7}},\ \bibinfo
  {pages} {A56} (\bibinfo {year} {2019}{\natexlab{b}})}\BibitemShut {NoStop}%
\bibitem [{\citenamefont {Okamura}(2020)}]{okamura2020linear}%
  \BibitemOpen
  \bibfield  {author} {\bibinfo {author} {\bibfnamefont {K.}~\bibnamefont
  {Okamura}},\ }\href@noop {} {\bibinfo {title} {Linear position measurements
  with minimum error-disturbance in each minimum uncertainty state}} (\bibinfo
  {year} {2020}),\ \Eprint {https://arxiv.org/abs/2012.12707} {arXiv:2012.12707
  [quant-ph]} \BibitemShut {NoStop}%
\bibitem [{\citenamefont {Ozawa}(1990)}]{ozawa1990quantum}%
  \BibitemOpen
  \bibfield  {author} {\bibinfo {author} {\bibfnamefont {M.}~\bibnamefont
  {Ozawa}},\ }\bibfield  {title} {\bibinfo {title} {Quantum-mechanical models
  of position measurements},\ }\href@noop {} {\bibfield  {journal} {\bibinfo
  {journal} {Phys. Rev. A}\ }\textbf {\bibinfo {volume} {41}},\ \bibinfo
  {pages} {1735} (\bibinfo {year} {1990})}\BibitemShut {NoStop}%
\bibitem [{\citenamefont {Ozawa}(1988)}]{ozawa1988measurement}%
  \BibitemOpen
  \bibfield  {author} {\bibinfo {author} {\bibfnamefont {M.}~\bibnamefont
  {Ozawa}},\ }\bibfield  {title} {\bibinfo {title} {Measurement breaking the
  standard quantum limit for free-mass position},\ }\href@noop {} {\bibfield
  {journal} {\bibinfo  {journal} {Phys. Rev. Lett.}\ }\textbf {\bibinfo
  {volume} {60}},\ \bibinfo {pages} {385} (\bibinfo {year} {1988})}\BibitemShut
  {NoStop}%
\bibitem [{\citenamefont {Caves}\ \emph {et~al.}(1980)\citenamefont {Caves},
  \citenamefont {Thorne}, \citenamefont {Drever}, \citenamefont {Sandberg},\
  and\ \citenamefont {Zimmermann}}]{caves1980measurement}%
  \BibitemOpen
  \bibfield  {author} {\bibinfo {author} {\bibfnamefont {C.~M.}\ \bibnamefont
  {Caves}}, \bibinfo {author} {\bibfnamefont {K.~S.}\ \bibnamefont {Thorne}},
  \bibinfo {author} {\bibfnamefont {R.~W.}\ \bibnamefont {Drever}}, \bibinfo
  {author} {\bibfnamefont {V.~D.}\ \bibnamefont {Sandberg}},\ and\ \bibinfo
  {author} {\bibfnamefont {M.}~\bibnamefont {Zimmermann}},\ }\bibfield  {title}
  {\bibinfo {title} {On the measurement of a weak classical force coupled to a
  quantum-mechanical oscillator. \text{I}. \text{I}ssues of principle},\
  }\href@noop {} {\bibfield  {journal} {\bibinfo  {journal} {Rev. Mod. Phys.}\
  }\textbf {\bibinfo {volume} {52}},\ \bibinfo {pages} {341} (\bibinfo {year}
  {1980})}\BibitemShut {NoStop}%
\bibitem [{\citenamefont {Yuen}(1983)}]{yuen1983contractive}%
  \BibitemOpen
  \bibfield  {author} {\bibinfo {author} {\bibfnamefont {H.~P.}\ \bibnamefont
  {Yuen}},\ }\bibfield  {title} {\bibinfo {title} {Contractive states and the
  standard quantum limit for monitoring free-mass positions},\ }\href@noop {}
  {\bibfield  {journal} {\bibinfo  {journal} {Phys. Rev. Lett.}\ }\textbf
  {\bibinfo {volume} {51}},\ \bibinfo {pages} {719} (\bibinfo {year}
  {1983})}\BibitemShut {NoStop}%
\bibitem [{\citenamefont {Caves}(1985)}]{caves1985defense}%
  \BibitemOpen
  \bibfield  {author} {\bibinfo {author} {\bibfnamefont {C.~M.}\ \bibnamefont
  {Caves}},\ }\bibfield  {title} {\bibinfo {title} {Defense of the standard
  quantum limit for free-mass position},\ }\href@noop {} {\bibfield  {journal}
  {\bibinfo  {journal} {Phys. Rev. Lett.}\ }\textbf {\bibinfo {volume} {54}},\
  \bibinfo {pages} {2465} (\bibinfo {year} {1985})}\BibitemShut {NoStop}%
\bibitem [{\citenamefont {Ozawa}(1989)}]{ozawa1989realization}%
  \BibitemOpen
  \bibfield  {author} {\bibinfo {author} {\bibfnamefont {M.}~\bibnamefont
  {Ozawa}},\ }\bibfield  {title} {\bibinfo {title} {Realization of measurement
  and the standard quantum limit},\ }in\ \href@noop {} {\emph {\bibinfo
  {booktitle} {Squeezed and Nonclassical Light}}}\ (\bibinfo  {publisher}
  {Springer},\ \bibinfo {year} {1989})\ pp.\ \bibinfo {pages}
  {263--286}\BibitemShut {NoStop}%
\bibitem [{\citenamefont {Maddox}(1988)}]{maddox1988beating}%
  \BibitemOpen
  \bibfield  {author} {\bibinfo {author} {\bibfnamefont {J.}~\bibnamefont
  {Maddox}},\ }\bibfield  {title} {\bibinfo {title} {Beating the quantum limits
  (cont'd)},\ }\href@noop {} {\bibfield  {journal} {\bibinfo  {journal}
  {Nature}\ }\textbf {\bibinfo {volume} {331}},\ \bibinfo {pages} {559}
  (\bibinfo {year} {1988})}\BibitemShut {NoStop}%
\bibitem [{\citenamefont {Arthurs}\ and\ \citenamefont
  {Kelly}(1965)}]{arthurs1965bstj}%
  \BibitemOpen
  \bibfield  {author} {\bibinfo {author} {\bibfnamefont {E.}~\bibnamefont
  {Arthurs}}\ and\ \bibinfo {author} {\bibfnamefont {J.}~\bibnamefont
  {Kelly}},\ }\bibfield  {title} {\bibinfo {title} {On the simultaneous
  measurement of a pair of conjugate observables},\ }\href@noop {} {\bibfield
  {journal} {\bibinfo  {journal} {The Bell Sys. Tech. J.}\ }\textbf {\bibinfo
  {volume} {44}},\ \bibinfo {pages} {725} (\bibinfo {year} {1965})}\BibitemShut
  {NoStop}%
\bibitem [{\citenamefont {Gauss}(1821)}]{Gauss1995theory}%
  \BibitemOpen
  \bibfield  {author} {\bibinfo {author} {\bibfnamefont {C.~F.}\ \bibnamefont
  {Gauss}},\ }\bibfield  {title} {\bibinfo {title} {Theoria combinationis
  observationum erroribus minimis obnoxiae, pars prior},\ }in\ \href
  {https://doi.org/10.1137/1.9781611971248} {\emph {\bibinfo {booktitle}
  {Commentationes Societatis Regiae Scientiarum Gottingensis Recentiores V
  (Classis Mathematicae)}}}\ (\bibinfo  {publisher} {societati regiae
  exhibita},\ \bibinfo {year} {febr. 15, 1821})\ \bibinfo {note}
  {\text{E}nglish translation: \textit{Theory of the Combination of
  Observations Least Subject to Errors, Part One, Part Two, Supplement},
  translated by G.W. Stewart (SIAM, Philadelphia, PA, 1995)},\ \Eprint
  {https://arxiv.org/abs/https://epubs.siam.org/doi/pdf/10.1137/1.9781611971248}
  {https://epubs.siam.org/doi/pdf/10.1137/1.9781611971248} \BibitemShut
  {NoStop}%
\bibitem [{\citenamefont {Ozawa}(1985)}]{ozawa1985}%
  \BibitemOpen
  \bibfield  {author} {\bibinfo {author} {\bibfnamefont {M.}~\bibnamefont
  {Ozawa}},\ }\bibfield  {title} {\bibinfo {title} {Conditional probability and
  a posteriori states in quantum mechanics},\ }\href
  {https://doi.org/10.2977/prims/1195179625} {\bibfield  {journal} {\bibinfo
  {journal} {Publ. Res. Inst. Math. Sci.}\ }\textbf {\bibinfo {volume} {21}},\
  \bibinfo {pages} {279} (\bibinfo {year} {1985})}\BibitemShut {NoStop}%
\bibitem [{\citenamefont {Okamura}\ and\ \citenamefont
  {Ozawa}(2016)}]{okamura2016measurement}%
  \BibitemOpen
  \bibfield  {author} {\bibinfo {author} {\bibfnamefont {K.}~\bibnamefont
  {Okamura}}\ and\ \bibinfo {author} {\bibfnamefont {M.}~\bibnamefont
  {Ozawa}},\ }\bibfield  {title} {\bibinfo {title} {Measurement theory in local
  quantum physics},\ }\href@noop {} {\bibfield  {journal} {\bibinfo  {journal}
  {J. Math. Phys.}\ }\textbf {\bibinfo {volume} {57}},\ \bibinfo {pages}
  {015209} (\bibinfo {year} {2016})}\BibitemShut {NoStop}%
\bibitem [{\citenamefont {Ozawa}(2019)}]{ozawa2019soundness}%
  \BibitemOpen
  \bibfield  {author} {\bibinfo {author} {\bibfnamefont {M.}~\bibnamefont
  {Ozawa}},\ }\bibfield  {title} {\bibinfo {title} {Soundness and completeness
  of quantum root-mean-square errors},\ }\href@noop {} {\bibfield  {journal}
  {\bibinfo  {journal} {npj Quantum Inf.}\ }\textbf {\bibinfo {volume} {5}},\
  \bibinfo {pages} {1} (\bibinfo {year} {2019})}\BibitemShut {NoStop}%
\end{thebibliography}%


\end{document}